\documentclass{article}
\usepackage{PRIMEarxiv}
\usepackage{amsmath,amsfonts}
\usepackage{algorithmic}
\usepackage{algorithm}
\usepackage{array}
\usepackage[caption=false,font=normalsize,labelfont=sf,textfont=sf]{subfig}
\usepackage{textcomp}
\usepackage{stfloats}
\usepackage{url}
\usepackage{verbatim}
\usepackage{color}
\usepackage{graphicx}
\usepackage{cite}
\usepackage{cuted} 
\usepackage{hyperref}

\usepackage{booktabs}
\usepackage{multirow} 
\usepackage{xcolor}
\newcommand{\bfw}[1]{\textbf{\textcolor{teal}{#1}}}

\begin{document}

\pagestyle{fancy}
\thispagestyle{empty}
\rhead{ \textit{ }} 
\title{CoDiNG -- Naming Game\\with Continuous Latent Opinions of Individual Agents}

\author{
	Mateusz Nurek \\
	Wroclaw University of Science and Technology \\
	Poland \\
	\texttt{mateusz.nurek@pwr.edu.pl} \\
	\And
	Joanna Ko{\l}aczek \\
	Wroclaw University of Science and Technology \\
	Poland \\
	\And
	Rados{\l}aw Michalski \\
	Wroclaw University of Science and Technology \\
	Poland \\
	\And
	Boleslaw K. Szymanski \\
	Rensselaer Polytechnic Institut \\
	NY, USA \\
	\And
	Omar Lizardo \\
	University of California Los Angeles \\
	CA, USA \\
}





\begin{center}
\parbox{0.9\textwidth}{
	\small
	\textit{This work has been submitted to the IEEE for possible publication. Copyright may be transferred without notice, after which this version may no longer be accessible.}
}
\end{center}

\maketitle

\begin{abstract}
Understanding the mechanisms behind opinion formation is crucial for gaining insight into the processes that shape the spread of political beliefs, cultural attitudes, consumer choices, and social movements in society. This work introduces a realistic model of opinion dynamics that captures the intricacies of real-world opinion dynamics by synthesizing principles from cognitive science. The proposed model is a hybrid continuous-discrete extension of the well-known Naming Game opinion model. The continuous layer captures the strength of each opinion through reinforcement and forgetting in the human brain, akin to memory imprints. The discrete layer allows for converting intrinsic continuous opinion into a discrete form, which often occurs when we publicly verbalize our opinions. We evaluated our model on longitudinal data combining real communication events with repeated surveys of the same individuals, comparing it against the Naming Game, the hybrid SJBO model, and four simple baselines at the population and the individual level. Unlike rigid baselines and the classic Naming Game, which inherently capture only a single aspect, hybrid models can be tuned to model either individual-level opinions or aggregate opinion dynamics. However, this flexibility comes with a strict trade-off, as they cannot accurately reproduce both simultaneously. Out of the six analysed topics, our model exceeds or matches SJBO, showing that reinforcement and forgetting grounded in cognition contribute to explaining opinion dynamics. Additionally, in our empirical data, individuals change their opinions while the aggregate distribution stays almost stationary, so a model reproducing no dynamics can still score well. This observation indicates that evaluating opinion models must be multidimensional and rely on more than one metric.
\end{abstract}


\section{Introduction}
As social animals, humans prefer living in groups rather than alone~\cite{dunbar2007evolution}. This style of living fosters frequent social interactions, directly between two individuals communicating and exchanging information and opinions about each other and indirectly when they discuss their acquaintances. These types of interactions are fundamental for spreading opinions, the diffusion of innovations, and influence in society, usually starting from the dyadic level~\cite{zbieg2012studying}, through small and large groups~\cite{friedkin2011social, centola2015social}, sometimes even reaching the whole population~\cite{granovetter1978threshold}. Classical sociological studies suggest that our interactions also build a platform where people either achieve a consensus, polarize, or are between these states~\cite{tarde1899}. One of the exciting research areas in more recent work in computational social science focuses on social interactions and aims to generate formal mathematical models of the spread of information, ideas, and influence in networked social systems~\cite{pierce2001global, rogers1994diffusion, friedkin1990social}.
 
A popular approach to modeling the auto-correlation of opinions among actors embedded in an interactive system uses geospatial models to estimate the total effect of diffusion or a final opinion type; typically these models see influence as produced by contiguity and propinquity. These include the parametric logistic model~\cite{carrington2005models}, the Bass model~\cite{bass1994bass}, or Moran's Spatial Autocorrelation Model~\cite{moran1950notes}. However, for the last three decades, a novel approach to opinion and influence dynamics beyond those generated by physical contiguity in space and more specifically grounded on complex networks has become dominant~\cite{friedkin1997social, guilbeault2018complex}. These social interaction-based models are useful when we are interested not only in the final spread or opinion distribution but also in how the process unfolds over time and across communities of connected individuals. This approach assumes either an explicit or implicit network of dyadic relations between actors plays the role of a transport layer (a medium) for the spread~\cite{atkin1977combinatorial}.
 
From the complex networks perspective, modeling of diffusion requires some underlying models of the relevant connectivity structure between actors, such as chain, lattice, circle, or more realistic ones, such as the Barabasi-Albert~\cite{albert2002statistical} Preferential Attachment model (PA) or the Watts-Strogatz~\cite{watts1998collective} Small World model (SW). When looking at real-world examples, including when analyzing genuine opinion spread, it is often overlooked that in most settings, the diffusing entity usually shapes the underlying network, so there is a complex interplay between these two layers that should not be ignored~\cite{karsai2011small, bahulkar2017coevolution}.

While internal opinions are often modeled as continuous variables, human communication frequently requires the verbalization of these stances into discrete choices. A fundamental limitation of traditional models dealing with such discrete opinion states, like the Naming Game, is their reliance on instantaneous and memoryless representations of social interactions. However, human memory and cognitive processes operate differently. Human perception of events changes continuously over time, and the strength of a memory trace gradually decays. We previously proposed the CogSNet model to capture these exact memory dynamics in temporal networks~\cite{michalski2021social}. Building on that foundation, we hypothesize that the same cognitive principles are crucial for realistic opinion modeling. Motivated by the findings from CogSNet, this work translates the mechanisms of human memory and forgetting into the domain of opinion dynamics.
 
In this work, we propose a new model called CoDiNG (\textbf{Co}ntinuous-\textbf{Di}screte \textbf{N}aming \textbf{G}ame) that extends the classic Naming Game model for opinion dynamics~\cite{baronchelli2006sharp, steels1995self} by introducing cognitive mechanisms aimed at modeling how an internalized attitude or commitment to a particular opinion is learned and subsequently decays in human memory if not subject to reinforcement. We characterize this cognitive aspect of opinion dynamics in networks using the CogSNet model~\cite{michalski2021social}, an innovative model of the temporal dynamics of learning and forgetting that was shown to outperform other non-cognitive approaches in modeling human memory imprints. The rationale behind the CogSNet model is rooted in the following social-scientific considerations:
 
We start with the well-established social psychological principle that publicly verbalizing an opinion entails discretizing a more complex internal state. As the sociologist C. Wright Mills put it in a classic statement on the role of language in social communication~\cite{mills1940situated}:

\begin{quote}
	Accompanying such quests for something more realistic and backed by rationalization is the view many sociologists hold that language is an external manifestation or concomitant of something prior, more genuine, and `deep' in the individual. `Real attitudes' versus `mere verbalization' or `opinion' implies that, at best, we only infer from his language what `really' is the individual's attitude or motive.
\end{quote}
 
Based on this and further work drawing on this principle~\cite{zaller1992nature}, we infer that the attitude an individual has at the point of verbalization needs to be converted into a more coherent linguistic token~\cite{De_Saussure1964-pz}. As such, even if a person wanted to express their opinion more subtly, incorporating all the nuances, they would have to first throw an anchor rooted in a discrete opinion set, and later, this person could elaborate more on these aspects. For instance, in the United States, if a person is asked about their political views and sees themselves as a Democrat but has some divergence from the set of values Democrats share, that person will first choose ``Democrat'' from their vocabulary and only afterward, will they make some corrections to align their verbalized opinion with the requisite internal state. To accommodate this dynamic, our proposed model incorporates two layers: The first, an internal one that is continuous and more complex, and the second, an external one that one achieves through verbalization. The external layer, rooted in the language, is discrete and allows for more accessible communication of opinions, even if it occasionally seems too simplistic compared to the internal state.
 
The proposed model's second pillar focuses on memory, a crucial element of human cognition~\cite{squire2004memory}. The quality of modeling of human memory becomes particularly significant when individuals need to retrieve extensive knowledge and select information from a noisy environment~\cite{neisser2000memory}. Consequently, one of the primary challenges in cognitive science has been comprehending the mechanisms involved in managing information in human memory~\cite{baddeley2004psychology, Mace2019-gl}, mainly when dealing with information overload~\cite{holyst2024protect}. Due to the highly complex nature of the human brain and the variations across individuals, establishing precise details of these mechanisms is challenging~\cite{huttenlocher2009neural}. Nonetheless, researchers in psychology and cognitive science have made significant progress in developing reliable working models of memory and other essential cognitive processes. For instance, the ACT-R cognitive architecture~\cite{anderson1997act} effectively models critical aspects of human declarative memory, successfully reproducing numerous well-known laboratory effects. Among these empirical regularities, the primacy and recency effects in list memory ~\cite{anderson1998integrated} are particularly significant for our purposes.
 
We have built the foundation for the CoDiNG model following the conclusions drawn from these two aspects of human social and cognitive functioning. We tested the proposed model on a real-world longitudinal dataset tracking individual opinion trajectories. We demonstrated that, unlike the classical Naming Game~\cite{baronchelli2006sharp}, our hybrid approach provides the flexibility needed to capture either individual-level opinion changes or macroscopic aggregate distributions. While navigating the inherent trade-off between these two scales, CoDiNG matches or outperforms alternative approaches, such as the SJBO model~\cite{fan2015emergence}. As such, CoDiNG joins the family of hybrid models of opinion formation, but its intrinsic properties make it more suitable for modeling processes where individuals hold conflicting or multiple opinions with differing levels of commitment and strength over time.

In summary, this article makes the following contributions:
\begin{enumerate}
    \item CoDiNG, a new hybrid continuous-discrete extension of the Naming Game in which each agent holds a separate internal weight for each opinion that is strengthened through interaction and forgotten over time.
    \item An evaluation of the model on real-world longitudinal data that combines communication events with surveys of the same individuals and compares CoDiNG with the classic Naming Game and the SJBO model as well as four simple baselines on six topics.
    \item The observation that opinion models cannot be assessed with a single metric, because in our data agents change their opinions while the aggregate distribution remains stable, so that a model which reproduces no dynamics at all can still achieve the best score.
\end{enumerate}
 
This work is organized as follows. In the next section, we present related research on opinion dynamics models. Next, in Section~\ref{sec:preliminaries}, we present the background of the classic Naming Game, which serves as the foundation for our work. We then introduce the proposed CoDiNG model in Section~\ref{sec:coding}, where we detail the integration of cognitive mechanisms of reinforcement and forgetting. Finally, we present the experimental setting and results to conclude the work and present future work directions in Section~\ref{sec:conclusions}.

\section{Related Research}
We divided this section into three parts. The first part discusses \textit{discrete} models where opinions typically take one of two values (e.g., agree and disagree); with these models, binary-state dynamics serve as foundational models for capturing essential aspects of social interaction. The idea of nodes switching between two states, influenced by their neighbors, provides a simplified yet revealing representation of how opinions evolve within a network. These models have been successfully applied to various scenarios, ranging from competition between opinions to the spread of information, rumors, or behavior~\cite{gleeson2013binary}. The second part discusses \textit{continuous} opinion models where opinions can take any value within a specified range, representing the strength of conviction towards a particular opinion. The third part discusses \textit{hybrid} models, which combine features of both discrete and continuous approaches.
 
\subsection{Discrete Opinion Models}
Models of opinion dynamics have a strong connection to physical models, where scientists have successfully utilized physical models for simulating opinion dynamics. An example of such a model that has gained significant popularity is the Ising model~\cite{ising1925beitrag}, initially designed for modeling phase transitions in ferromagnets. Still, its mechanics have also found reflection in opinion modeling~\cite{fortunato2013statistical}. In its original form, the Sznajd model~\cite{sznajd2000opinion} is a binary opinion model that extends the Ising model by incorporating a social validation mechanism. If a pair of randomly selected nodes \(v_i\) and \(v_j\) have the same opinion, they impose the same opinion on their neighbors. If the selected pair has opposite opinions, \(v_i\)'s neighbors change their opinion to the opposite of \(v_i\)'s opinion, and \(v_j\)'s neighbors change their opinion to the opposite of \(v_j\)'s opinion. Alternatively, if the selected pair has contradictory opinions, we can use another mechanism -- do nothing. The Sznajd Model was initially tested on a 1-D lattice and later extended to the square lattice~\cite{stauffer2000generalization, bernardes2001damage} as well as various networks~\cite{bernardes2002election, elgazzar2001application, ru2008opinion}.
 
Other common generalizations of binary-state dynamics are the Global Threshold Model (GTM)~\cite{granovetter1978threshold}, the Network Threshold Model (NTM)~\cite{valente1996social}, and the Independent Cascade Model (ICM)~\cite{goldenberg2001talk}. Granovetter's GTM is a deterministic model that utilizes the pressure from others, which, upon surpassing a predefined threshold, defined as the proportion of the population that has already adopted it, leads to a change in opinion. Valente's Network Threshold Model is like Granovetter's, except that everyone defines a threshold as a fraction of neighbors that have already adopted---within a specific network topology---rather than the entire population. ICM is a nondeterministic model that focuses on pairwise interactions and assumes that each neighbor has a certain probability of causing a change in the opinion of a given node in the network. Another well-known non deterministic model of opinion is the Voter Model~\cite{clifford1973model}. A random individual is selected, and that voter's opinion undergoes a transformation defined by the views of their neighbors. The specific mechanism involves choosing one of the selected voter's neighbors according to a given set of probabilities and then transferring that neighbor's opinion to the selected voter.
 
All the models discussed so far, except for Granovetter's, operate on the assumption that a network is constructed, allowing us to utilize information about each node's neighborhood. Building the graph and performing operations on large social networks can be time-consuming. Fortunately, literature provides solutions that can simulate opinion formation in a network without explicitly constructing it. For instance, the Galam model~\cite{galam1999application} involves randomly assigning nodes to groups of size \textit{r}, where all nodes within a group adopt the majority opinion. Following this, nodes randomly switch to new groups, repeating the entire opinion change process.
 
Another significant and well-researched discrete opinion model, which also serves as the foundation for our CoDiNG model, is the two-word version of the Naming Game~\cite{baronchelli2006sharp}. This model arises from studies in multi-agent systems~\cite{steels1995self} and employs statistical mechanics to interpret social dynamics~\cite{fortunato2013statistical}. Typically, this model utilizes a mean-field approximation in which any agent can interact with any other agent, although it also accommodates various network topologies~\cite{baronchelli2016gentle}. In this model, nodes share information about their opinions, and based on the received data, they modify their views accordingly. Apart from the presence of one of two opinions (A or B), a vital feature of this model is that a node can also hold a mixed opinion (AB). This aspect provides a closer representation of real-world scenarios where individuals, in addition to having distinct opinions (A or B), may experience uncertainty about their views, leading them to adopt a mixed opinion (AB). Sociologists call this phenomenon "cross-pressure"~\cite{davis1963structural}.
 
\subsection{Continuous Opinion Models}
The DeGroot opinion model~\cite{degroot1974reaching} is a classic approach to opinion dynamics, where individuals update their opinions iteratively by employing a weighted average of their neighbors' views. This process facilitates convergence toward a consensus over time. The model captures the idea that individuals adjust their beliefs through social influence~\cite{friedkin1990social, friedkin1997social, friedkin2011social}, highlighting the role of connected peers in shaping collective opinions.
 
Among continuous opinion models, we can identify a group of bounded confidence models, which includes the Deffuant-Weisbuch model (DW)~\cite{deffuant2000mixing} and the Hegselmann-Krause model (HK)~\cite{rainer2002opinion}. The DW model incorporates bounded confidence by randomly selecting two agents and allowing them to adjust their opinions only if their difference is less than a predefined threshold, $\delta$. This setup limits social influence in cases where individuals have a sufficiently high degree of similarity in their opinions.
Among continuous models, the DW model is the only one whose operation can be described at the level of events, eliminating the need to construct the entire network and gather information about neighbors. The HK model for each node synchronously updates the opinions of all neighbors whose beliefs differ from the opinion of this node less than the predefined threshold. This mechanism mimics the real-world impact of a group on an individual.
 
\subsection{Hybrid Models}
The third least numerous group of opinion dynamics models consists of hybrid models, incorporating elements of both discrete and continuous models. This group includes the CODA model, which extends the Voter and Sznajd Models~\cite{martins2008continuous}. The general assumption of this model is that agents exhibit discrete actions externally (preference for one of two opinions) while internally possessing a continuous opinion expressing the strength of conviction. Inspired by the CODA model, Zhan et al.~\cite{zhan2021bounded} proposed the SNOAE model, where connected nodes in the network can directly observe the continuous opinions of their neighbors. In contrast, unconnected nodes only see the discrete actions of other nodes.
 
Finally, the SJBO model~\cite{fan2015emergence, fan2016opinion} extends the concept of discrete actions and continuous opinions to include a third state symbolizing uncertain opinion. Our model uses a similar approach to the SJBO model with two notable differences, distinguishing it from previous hybrid models. Firstly, CoDiNG implements an opinion decay mechanism firmly rooted in empirical research in the cognitive sciences, making it akin to how traces in memory are reinforced and forgotten in our minds. Secondly, our model can simultaneously compute the strength of conviction for both agents’ opinions. Such computation is impossible in other hybrid models since they measure two agents’ opinions on a single scale. Additionally, in contrast to CODA and SNOAE, our model allows agents to have mixed opinions, bringing them closer to the real world, where someone may be uncertain about their opinion.
\section{Preliminaries: Naming Game}
\label{sec:preliminaries}
In the two-opinion variant of the Naming Game~\cite{baronchelli2006sharp}, also known as the \textit{Binary Agreement Model}, nodes can hold either one of two conflicting opinions or both opinions simultaneously~\cite{xie2011social}. At each time step, a model randomly selects a speaker and also randomly chooses one of its neighbors as the listener. The speaker then communicates their opinion to the listener (randomly selected if the speaker holds two opinions). If the listener already holds the conveyed opinion, both the speaker and the listener retain it, eliminating all other opinions. Table~\ref{t2} shows a schema for this process.
            	
            	\begin{table}[H]
                            	\centering
                            	\caption{Interaction rules of the Naming Game. Each row gives the states of the sender--listener pair before and after an interaction for the opinion conveyed (arrow).}
                            	\begin{tabular}{lcc}
                                            	\hline
                                            	\textbf{Before Interaction} & \textbf{After Interaction} \\
                                            	\hline
                                            	$A \xrightarrow{A} A$ & $A - A$ \\
                                            	$A \xrightarrow{A} B$ & $A - AB$ \\
                                            	$A \xrightarrow{A} AB$ & $A - A$ \\
                                            	$B \xrightarrow{B} A$ & $B - AB$ \\
                                            	$B \xrightarrow{B} B$ & $B - B$ \\
                                            	$B \xrightarrow{B} AB$ & $B - B$ \\
                                            	$AB \xrightarrow{A} A$ & $A - A$ \\
                                            	$AB \xrightarrow{A} B$ & $AB - AB$ \\
                                            	$AB \xrightarrow{A} AB$ & $A - A$ \\
                                            	$AB \xrightarrow{B} A$ & $AB - AB$ \\
                                            	$AB \xrightarrow{B} B$ & $B - B$ \\
                                            	$AB \xrightarrow{B} AB$ & $B - B$ \\
                                            	\hline
                            	\end{tabular}
                            	\label{t2}
            	\end{table}

The Naming Game differs from other opinion dynamics models by allowing agents to hold both opinions simultaneously. This feature reduces the time required to achieve consensus from uniform initial conditions. The model can implement various scenarios, including those cases in which changes in state are not deliberate or are randomly
calculated to mimic unconsciousness. It can also show how opinions evolve when individuals adapt their views through social interactions~\cite{xie2011social, xie2012evolution}.

\section{CoDiNG - New Model of Opinions with Latent States}
\label{sec:coding}

In this work we propose the Continuous-Discrete Naming Game (CoDiNG) which combines the classic discrete Naming Game model with an additional continuous layer. The model adapts the cognitive principles of memory and forgetting introduced in our previous work on Cognitive Social Networks (CogSNet)~\cite{michalski2021social} and integrates them directly into the opinion dynamics. The continuous layer models the strength of agent internal beliefs and the discrete layer translates it into the opinion the agent publicly expresses.

\subsection{Agent State and Opinion Translation}
\label{subsec:agent_state}

Each agent $i$ has two independent continuous variables $w_{i,A}(t)$ and $w_{i,B}(t)$ representing the internal weights of their belief in opinion $A$ and opinion $B$ at time $t$. These weights fall within the range $[0, 1]$.

In CoDiNG, opinion dynamics is driven by agent interactions. During interactions, agents observe only the discrete states $o_i(t) \in \{A, B, AB\}$ where $AB$ represents a neutral state. The translation from the continuous latent state to the discrete verbalized opinion is governed by a threshold parameter $\gamma \in (0,1)$. For any agent $i$ at time $t$ the discrete opinion is determined by the absolute difference between their internal weights $\Delta_o = |w_{i,A}(t) - w_{i,B}(t)|$:

\begin{equation}
\label{coding_discrete}
	o_i(t) = 
	\begin{cases} 
		A & \text{if } \Delta_o \geq \gamma \text{ and } w_{i,A}(t) > w_{i,B}(t) \\ 
		B & \text{if } \Delta_o \geq \gamma \text{ and } w_{i,A}(t) < w_{i,B}(t) \\ 
		AB & \text{if } \Delta_o < \gamma 
	\end{cases}
\end{equation}

\subsection{Memory Mechanism}
\label{subsec:memory}

CoDiNG is inspired by the memory mechanisms of the human mind. Therefore, similar to how memory imprints can be reinforced or weakened, opinions are strengthened through frequent interaction and forgotten over time if no interactions occur. The weight of a specific opinion $x \in \{A, B\}$ at time $t$ for agent $i$ is calculated as follows:
\begin{equation}
	w_{i,x}(t) = w_{i,x}(t_{i,x}) \cdot f(\Delta t)
	\label{eq:forgetting_general}
\end{equation}
where $f$ is the forgetting function and $\Delta t = t - t_{i,x}$ is the time elapsed since the last interaction of agent $i$ with opinion $x$ up to the current time $t$. The forgetting function is a monotonically non-increasing function of time, with $f(0)$ equal to 1 and $f(t) \geq 0$ for all $t > 0$. In our study we use two forgetting functions commonly observed in human memory modeling: the exponential function $f(\Delta t) = e^{-\lambda \Delta t}$ and the power-law function $f(\Delta t) = max(1, \Delta t)^{-\lambda}$.

Each interaction with opinion $A$ or $B$ increases the corresponding weight. The new weight is calculated by combining the reinforcement parameter $\mu \in [0,1]$ with the current decayed weight of that opinion. The update in the CoDiNG model is performed asynchronously according to the following update rule:
\begin{equation}
	w_{i,x}(t) = \mu + w_{i,x}(t_{i,x}) \cdot f(t-t_{i,x})  \cdot (1 - \mu)
	\label{eq:weight_update}
\end{equation}

\begin{figure}[ht]
	\begin{center}
        \includegraphics[scale=0.32]{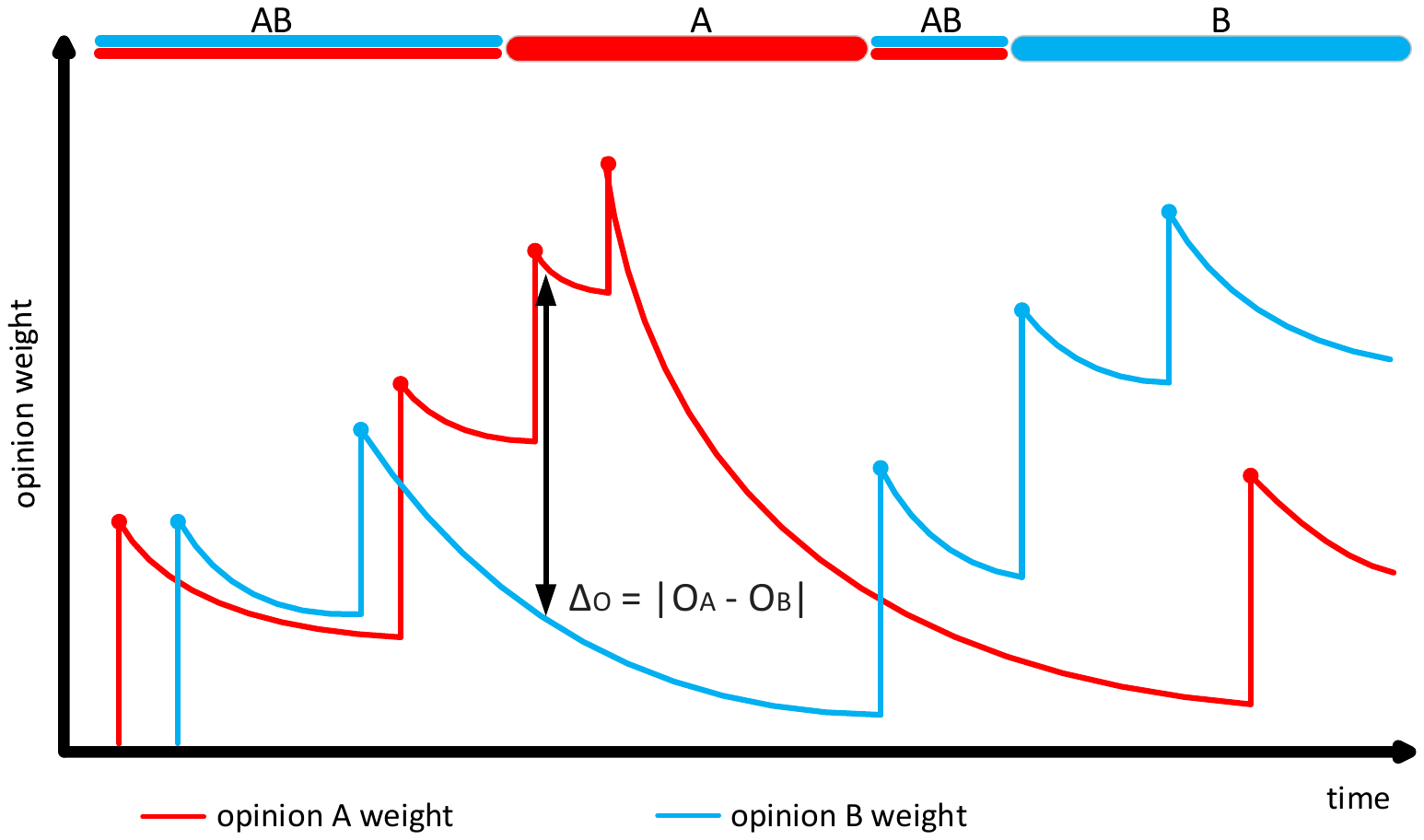}
        \caption{Illustration of opinion change in the CoDiNG model. The two curves show the internal weights of opinions A and B for a single agent over time. Each interaction reinforces the corresponding weight, while both weights decay between interactions according to the forgetting function. The expressed discrete opinion (top bar) follows from the weight gap $\Delta o = |o_A - o_B|$ relative to the threshold $\gamma$. The agent verbalizes A or B when $\Delta o \geq \gamma$ and stays in the neutral AB state otherwise.}
        \label{fig:coding-opinion}
	\end{center}
\end{figure}

The mechanics of the model is graphically presented in Figure~\ref{fig:coding-opinion}.

\subsection{Modified Naming Game Dynamics}

\label{subsec:dynamics}
The interaction mechanism in CoDiNG adapts the Naming Game rules to the continuous layer update. Each interaction strengthens the belief of both the recipient and the sender. This reflects social influence as well as the phenomenon where publicly expressing a stance reinforces commitment to it.

Suppose that at time $t$ a sender $s$ interacts with a recipient~$r$. The interaction proceeds in two steps. 

First, the sender $s$ selects an opinion to express based on Equation~\ref{coding_discrete}. If $o_s(t) = A$ or $B$, they express $A$ or $B$. If the sender is in the neutral state $o_s(t) = AB$, they randomly select either $A$ or $B$ proportionally to the internal weights $w_{s,A}(t)$ and $w_{s,B}(t)$.

Second, both the sender $s$ and the recipient $r$ reinforce their  internal weight for the expressed opinion according to Equation~\ref{eq:weight_update}.

\begin{figure}[]
	\begin{center}
        \includegraphics[scale=0.3]{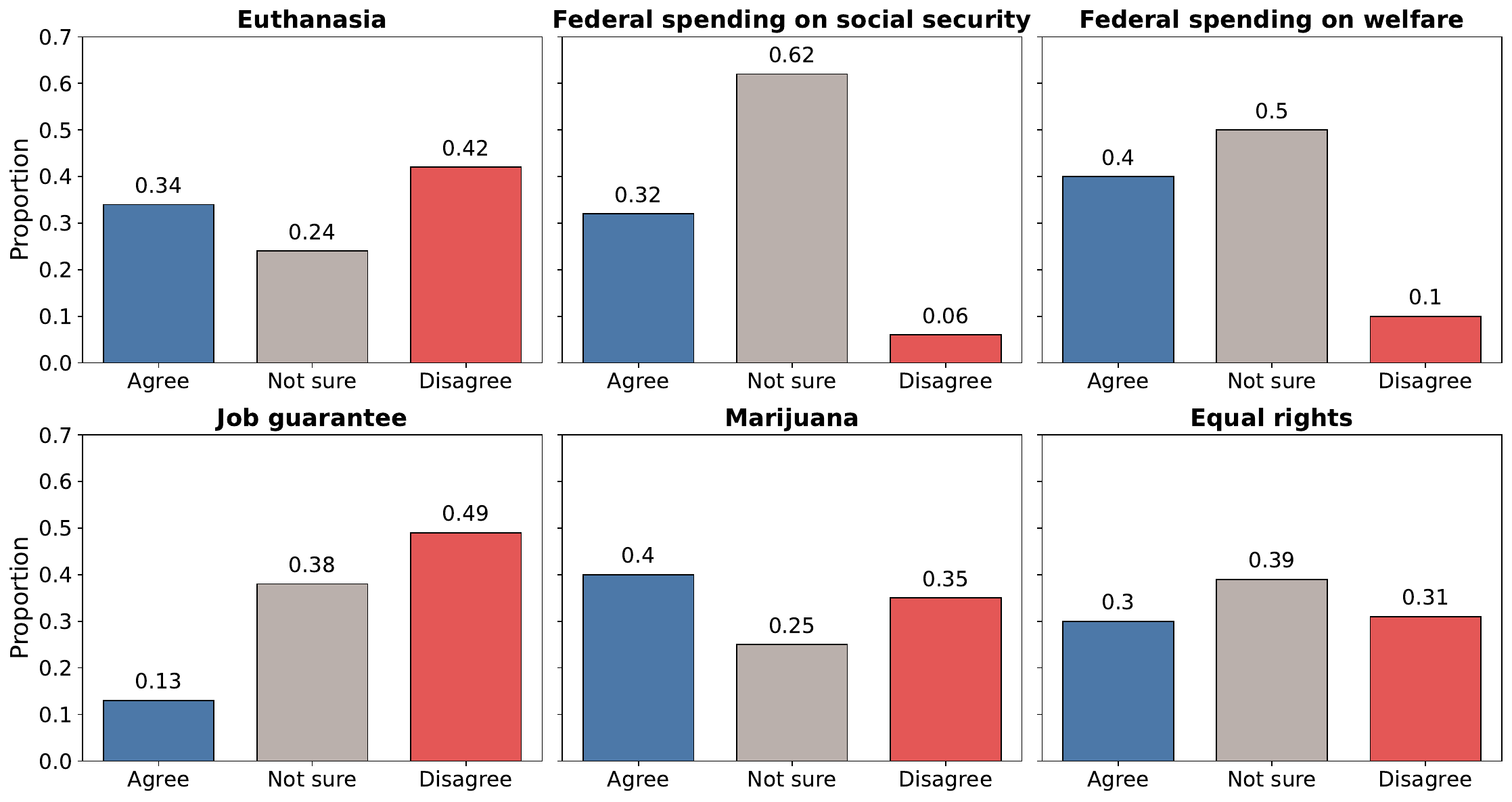}
        \caption{Distribution of survey answers for the six analyzed topics. Opinions concentrate in the neutral \textit{not sure} category for federal spending on social security, federal spending on welfare, and equal rights, indicating widespread uncertainty, whereas euthanasia, marijuana, and job guarantee are more polarized.}
    	\label{fig:dist}
	\end{center}
\end{figure}

\begin{figure}[]
	\begin{center}
        \includegraphics[scale=0.37]{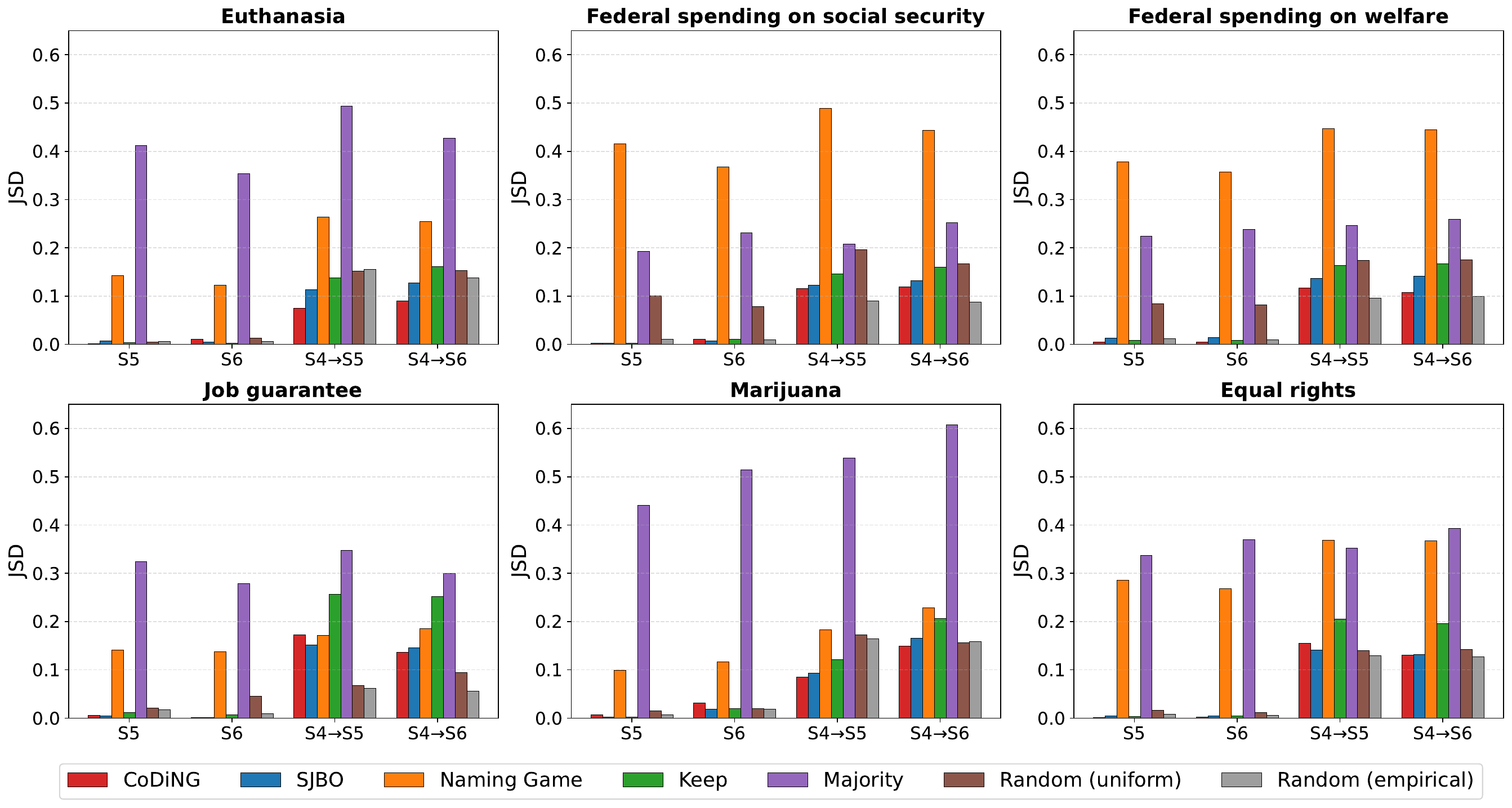}
        \caption{Macro-level evaluation. Jensen--Shannon divergence (JSD; lower is better) between the simulated and empirical opinion distributions, shown per topic for the compared models. Within each panel, the first two groups (S5, S6) report the static opinion distribution JSD at surveys five and six, and the last two (S4$\rightarrow$S5, S4$\rightarrow$S6) report the opinion trajectory JSD, (the divergence between the simulated and empirical distributions of opinion transitions from survey four).}
        \label{fig:jsd}
	\end{center}
\end{figure}

\begin{figure}[]
	\begin{center}
        \includegraphics[scale=0.37]{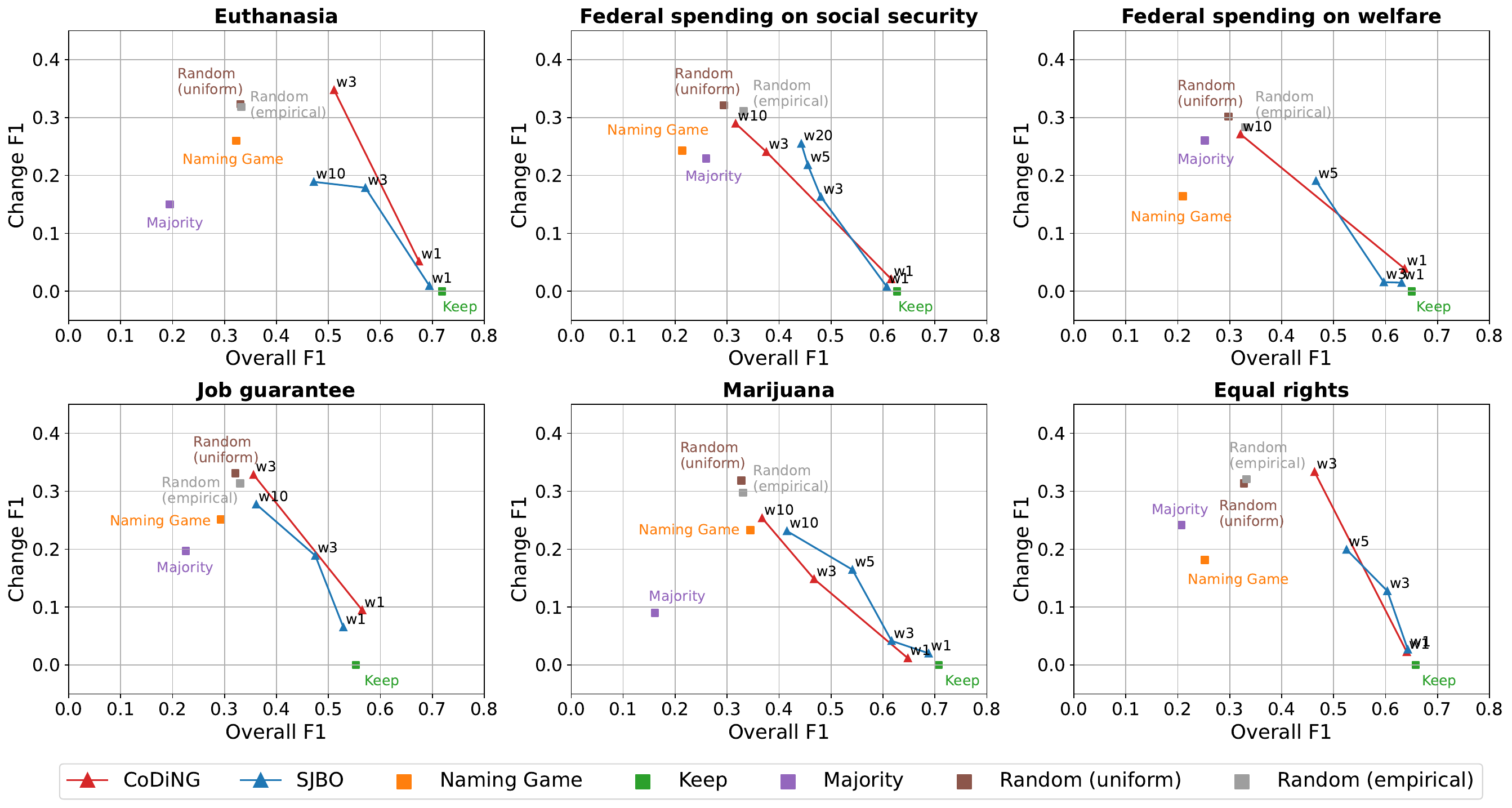}
        \caption{Micro-level evaluation. Trade-off between the overall F1 score ($x$-axis) and the change F1 score, computed only over agents who change opinion ($y$-axis), shown per topic. For CoDiNG and SJBO, the connected markers trace a Pareto front obtained by increasing the sample weight assigned to opinion-change cases during training (labeled $w_k$, where $k$ is assigned weight). The remaining models appear as single points.}
        \label{fig:pareto}
	\end{center}
\end{figure}

\section{Experiments}
\subsection{Dataset}
\label{subsec:dataset}
The data used to validate the model comes from \textit{NetSense} study~\cite{striegel2013lessons}, fielded from 2011 to 2013. The first part of the dataset contains information about communication events between students---sender ID, receiver ID, and timestamp. The second part of the dataset consists of social surveys conducted every semester in which students marked their opinions on various political and social topics. The Institutional Review Board (IRB) of the University of Notre Dame has reviewed the NetSense study and has approved it - the IRB Number is FWA 00002462. All participants provided written informed consent prior to taking part in the study.

Six topics were selected from the NetSense study surveys to evaluate our model based on preliminary analyses indicating which opinions exhibited the most change across consecutive semesters. We present the topics below.
 
\begin{enumerate}
	\item \textbf{Euthanasia}. When a person has an incurable disease, do you think doctors should be allowed by law to end the patient’s life by some painless means if the patient and his family request it?
	\item \textbf{Federal spending on social security}. Do you think federal spending on social security should increase, be kept the same, or decrease?
	\item \textbf{Federal spending on welfare}. Do you think federal spending on welfare should increase, be kept the same, or decrease?
	\item \textbf{Job guarantee}. Some people feel that the government should see to it that every person has a job and a good standard of living. Others think the government should just let each person get ahead on her/their own. Where would you place yourself on the 7-point scale?
	\item \textbf{Marijuana}. Do you think the use of marijuana should be made legal or not?
	\item \textbf{Equal rights}. Do you agree that we have pushed for equal rights in this country too far?
\end{enumerate}
 
The survey collected answers using different scales than the model, e.g., yes/no/not sure or a 7-level scale expressing the degree of agreement with the given question. The model can recognize three states, so we transformed all the above answers to the three-level scale: agree, disagree, and not sure. The transformed responses match a ternary pattern. The distributions of answers revealed that opinions tended to cluster around the middle, suggesting uncertainty on specific issues (federal spending on social security and welfare, equal rights). In contrast, the views on remaining topics exhibited polarization (euthanasia, marijuana, job guarantee) (Figure \ref{fig:dist}). While there was no notable correlation in how individuals responded to specific topics, the overall level was higher in the fifth survey than in the first~(Appendix C).

\begin{table}[]
    \centering
    \caption{Empirical opinion shift and aggregate distribution similarity between survey four and surveys five and six. The opinion shift is the fraction of agents whose discrete opinion changes relative to survey four. JSD is the Jensen--Shannon divergence between the two aggregate opinion distributions.}
    \label{tab:opinion_changes}
    \begin{tabular}{l c c c c}
        \toprule
        & \multicolumn{2}{c}{\textbf{Survey 4 $\rightarrow$ 5}} & \multicolumn{2}{c}{\textbf{Survey 4 $\rightarrow$ 6}} \\
        \cmidrule(lr){2-3} \cmidrule(lr){4-5}
        \textbf{Topic} & \textbf{Opinion Shift} & \textbf{JSD} & \textbf{Opinion Shift} & \textbf{JSD} \\
        \midrule
        euthanasia      & 0.25 & 0.004 & 0.28 & 0.003 \\
        fssocsec        & 0.26 & 0.003 & 0.28 & 0.010 \\
        fswelfare       & 0.28 & 0.008 & 0.29 & 0.009 \\
        jobguar         & 0.42 & 0.011 & 0.41 & 0.007 \\
        marijuana       & 0.22 & 0.002 & 0.34 & 0.019 \\
        toomucheqrights & 0.35 & 0.004 & 0.34 & 0.004 \\
        \bottomrule
    \end{tabular}
\end{table}

    
    
    
 
\subsection{Experimental Setting}
\label{subsec:setting}
To fit the parameters of CoDiNG, we split the interaction and survey data into training and testing (see Fig. D.1 in Appendix D). In the training phase, opinions from the first survey were used to initialize the agents' discrete and continuous states. Discrete states were set to one of three values (agree, not sure, or disagree) according to the survey answers, and the continuous states were computed from these discrete states. If the answer was agree or disagree, a high weight $(\geq 0.66)$ was set for the given opinion and a low weight $(\leq 0.33)$ for the opposite opinion. When the answer was not sure, we set the weights for both opinions between 0.33 and 0.66. Additionally, we checked whether the difference in the initialized weights satisfied the condition imposed by the tested $\gamma$ value. We ran a random search that tested 200 combinations of CoDiNG parameters (see Appendix A and B). We ran the simulation up to the time of survey four, and since each interaction has a timestamp, we saved the state of each agent at the completion time of surveys two, three, and four. During the simulation, the agent states were only saved, without any correction of the opinions from the consecutive surveys, so that we could measure the divergence of the model from the surveys over short and longer periods. During the training, every parameter combination was tested twenty times due to the randomness of the model, and the best combination was chosen based on the macro-averaged F1 score. The macro-averaged F1 score was computed separately at surveys two, three, and four and then averaged over these surveys and over all runs, to obtain parameters suited for both short- and long-term modeling of opinion dynamics.

Next, in the testing phase, we used the best combination of parameters and initialized the agents' discrete states based on the answers from survey four, and we assigned the weights in the continuous state in the same way as in the training phase. Answers from surveys five and six served as ground truth for final evaluation. We ran 100 simulations and averaged the macro F1 scores computed separately for each survey. We compared our model to simple baselines: keeping the same opinion, changing to the majority opinion, random change with uniform probability, and random change with probability proportional to the initial opinion distribution. Furthermore, we compared CoDiNG to the classic Naming Game and to the SJBO model. SJBO is an opinion model based on social judgment theory, with assimilation of similar opinions, repulsion of opposite ones, and neutrality toward differing opinions that remain below the repulsion threshold. Unlike the Naming Game, the SJBO model has parameters that require tuning. To ensure a fair comparison, we optimized them on the training data using the exact same protocol as described for CoDiNG. Similarly to CoDiNG, agents in SJBO can also have one of three discrete states as well as a continuous state, which allows us to directly compare the two models, one based on a cognitive mechanism and the other on social psychology.

Additionally, we performed a sensitivity analysis in which, starting from the best parameters for each topic, we varied a single parameter at a time while keeping the others fixed, to assess how sensitive the model is to parameter changes. We also tested the extreme allowed values, which include an ablation of the reinforcement and memory mechanisms by turning them off, to better understand their contribution and the model's behavior.

\subsection{Evaluation Metrics and Trade-Offs}
Evaluation of opinion models is not a trivial task, because they can be accurate in more than one sense. Opinion dynamics can be captured at a macro level, where the model precisely simulates the overall distribution of the system, or at a micro level, where we look at individual agents and their opinions. These two levels do not have to move together. In our data, individuals change their opinions substantially while the aggregate distribution barely moves. In Table~\ref{tab:opinion_changes}, we present the opinion change in the test part of the data. We can see that for most topics the opinion change of agents between surveys is 22--35\%, with the exception of the job guarantee topic, which exceeds 40\%. This indicates some dynamics at the micro level, however at the macro level comparing the distributions gives a Jensen--Shannon divergence (JSD) close to zero, which implies a stable distribution.

As mentioned in the previous section, we decided to use the macro-averaged F1 score to find the best parameters for the CoDiNG model, which is justified since only micro-level dynamics carry information about opinion change. However, this metric also has a disadvantage because the majority of people in the dataset do not change their opinion. It is possible to maximize F1 by choosing parameters that push the model toward keeping opinions unchanged. This is why, alongside the overall F1, we also report a change F1, computed only over the agents who changed their opinion, to assess how well the model captures opinion change. To provide a comprehensive macro-level evaluation, we also compute two variants of the JSD. The first, opinion distribution JSD, measures the divergence between the simulated aggregate opinion distribution and the empirical survey answers at a specific time step. The second, opinion trajectory JSD, evaluates system dynamics by comparing the distribution of opinion transitions (e.g., shifts from one specific opinion to another between two surveys) against the actual empirical transitions.

To make the model more sensitive to opinion change, we also tested different sample weights for these cases while searching for the best parameters during the training phase. Cases in which an agent did not change opinion were always assigned a weight of 1, while for cases with an opinion change we tested the following weights: 3, 5, 10, 20, 40, and 80. We then constructed a Pareto front to show the trade-off between the overall F1 and the change F1 depending on the applied weight.

Statistical significance against baseline and competing models was assessed using a one-sided Wilcoxon signed-rank test on per-run metrics and effect size using rank-biserial correlation. To correct for multiple comparisons, the Benjamini--Hochberg FDR procedure was applied separately for each topic and within three distinct metric families: opinion distribution JSD, opinion trajectory JSD, and the overall F1 score.
 
\section{Results}

\subsection{Macro-Level Analysis}
We first assessed how accurately each model reproduces macro-level opinion dynamics. Figure~\ref{fig:jsd} shows the JSD between the simulated and empirical survey distributions. The distribution JSD measures how far the simulated distribution diverges from the empirical opinion distribution at a given evaluated survey, five (S5) and six (S6). The trajectory JSD measures the divergence between the simulated and empirical distributions of opinion transitions between the initial survey (S4) and the two subsequent surveys, five (S4$\to$S5) and six (S4$\to$S6).

Most models reproduce the real distribution well, achieving a low JSD for surveys five and six. CoDiNG, SJBO, Keep, and the random baselines are all close to zero on this metric. Because the opinion distribution is nearly stationary in our data (Table~\ref{tab:opinion_changes}), even simple baselines such as Keep or the random baselines can achieve a low divergence. The clear exception is the Naming Game and Majority baselines, which are the two worst models on every topic, reaching a high divergence. Statistical tests (Table~\ref{tab:stat_tests}) show that CoDiNG significantly outperforms the Naming Game and Majority on this metric in all six topics, with near-maximal effect sizes. For the remaining baselines, CoDiNG outperforms them in most topics at survey five, except for marijuana and in case of comparison with SJBO also job guarantee. For the longer horizon at survey six, the model's output diverges more from the empirical distribution. Consequently, CoDiNG no longer significantly outperforms SJBO, Keep, and the random baselines on the individual topics of euthanasia and marijuana, nor on the more social topic of federal spending on social security. As with survey five, job guarantee remains an additional exception in the comparison with SJBO. Notably, euthanasia, marijuana, and job guarantee are highly polarized topics, whereas federal spending on social security is dominated by uncertainty (Fig.~\ref{fig:dist}).

The trajectory JSD is far more discriminating and lets us evaluate how well the models capture opinion dynamics at the macro level, since it describes the distribution of transitions. CoDiNG significantly outperforms all baselines on this metric in at least four topics, except for the empirical-random baseline (Table~\ref{tab:stat_tests}). The advantage of CoDiNG is visible in its more accurate modeling of the opinion-change trajectory, especially for the individual topics. Compared with the empirical-random baseline, CoDiNG outperforms it only on the individual topics, euthanasia and marijuana. The Naming Game, Majority, and Keep baselines capture opinion dynamics the worst, reaching the highest trajectory JSD. Both hybrid models, CoDiNG and SJBO, capture the dynamics better than the discrete Naming Game; however, because the distribution is stable across surveys, changing opinions randomly in proportion to the initial opinion distribution remains a strong baseline.

\subsection{Micro-Level Analysis}
We now examine how well each model captures the opinions of individual agents, using the overall and change F1 scores shown as a Pareto front in Figure~\ref{fig:pareto}. To avoid cherry-picking, the statistical tests uses CoDiNG in its default configuration, with equal weighting of changed and unchanged cases, and evaluates the overall F1 alone (Table~\ref{tab:stat_tests}). The front over sample weights is reported to expose the trade-off inherent in modeling opinion dynamics, not to claim that one model is generally best.

Overall F1 can be maximized without reproducing any dynamics, therefore the Keep baseline reaches a high overall F1 with change F1 equal to zero. The random baselines show the opposite result. They achieve much lower overall F1 but reach the best change F1 among the compared models in four out of six topics. Only for euthanasia and equal rights was CoDiNG able to outperform them, and only with higher weights for changing opinion samples. The Majority and Naming Game baselines are not satisfactory on either metric.

Among the compared models, only CoDiNG and SJBO combine a relatively high overall F1 with a non-trivial change F1. On most topics their fronts overlap in the conservative region, where both leave most opinions unchanged, and neither has an advantage. They separate when higher sample weights push them toward detecting change, and even then the separation depends on the topic rather than holding on every one. On several topics the two fronts stay close along their whole length, which suggests that the two mechanisms behave similarly there. The clearest exception is euthanasia and equal rights, where CoDiNG reaches a distinctly higher change F1 while SJBO stops at a lower value and cannot be pushed further. The advantage of CoDiNG over SJBO at the micro level is therefore limited to a few topics rather than general, and at the macro level it is reinforced by its generally lower trajectory JSD.

\subsection{Sensitivity and Ablation Analysis}
\label{subsec:sensitivity}
We took the best configuration for each topic, froze it, and varied one parameter at a time to observe how the model reacts. The range of tested values reaches the extreme allowed values, which switch the reinforcement and the forgetting off, so the ablation is contained in the same analysis. The threshold $\gamma$ varies between $10^{-5}$ and 0.99, since the model needs three states and the values 0 and 1 would leave no room for the not sure state. We show federal spending on social security and job guarantee as representative topics, and the remaining topics are reported in the Appendix E.

The selected parameters are almost the same for every topic. The reinforcement peak $\mu$ equals 0.0001 everywhere. The forgetting intensity $\lambda$ equals $10^{-6}$ for five topics with power-law forgetting function and $10^{-7}$ for job guarantee with exponential forgetting. The threshold $\gamma$ equals 0.7 for five topics and drops to 0.03 only for job guarantee, which is also the topic with the largest number of opinion changes. The model therefore shares most of the parameter values regardless of the topic, which suggests a general range of values fitted to the mechanics of the model rather than overfitting to the data.

Setting the reinforcement to zero switches the mechanism off, however it does not reduce the model to the Keep baseline. The forgetting can still shrink the gap between the two opinion weights until an agent falls back to the not sure state. Increasing the reinforcement improves the change F1 and the trajectory JSD, but at the same time it deteriorates the overall F1. For social security the trajectory JSD starts to increase above a certain range while the change F1 flattens.

When $\lambda$ is set to zero, CoDiNG has a permanent memory in which weights grow toward saturation because nothing is forgotten. For both topics all metrics are almost the same for the best $\lambda$ and for $\lambda = 0$, so the forgetting mechanism is not crucial at the selected configuration. As $\lambda$ grows, faster forgetting lowers the overall F1. For job guarantee, with exponential forgetting, the deterioration is faster and more abrupt, whereas for social security, with power-law forgetting, it is more gradual. At the same time, a larger $\lambda$ improves the change F1 on both topics, which shows that forgetting contributes to detecting opinion change. The effect is stronger for social security, where the trajectory JSD improves as well, although too fast forgetting eventually raises the JSD and flattens the change F1. For job guarantee the improvement in the change F1 is not accompanied by a better trajectory JSD, and faster forgetting always makes it worse. In general, a shorter memory helps to model opinion change, because the agent relies more on recent influence, but the effect always depends also on the threshold $\gamma$. The decay erases the accumulated history that anchors an agent to its current opinion, while $\gamma$ sets how large the gap between the opinion weights must be before the agent switches, so faster forgetting and a lower threshold can make opinion change easier.

The threshold $\gamma$ is the most influential parameter of the model. It sets the gap that the two opinion weights must cross before an agent changes its state, and since the selected $\mu$ is much smaller than $\gamma$, crossing it requires a large number of consistent interactions. A small $\gamma$ makes agents choose one of the sharp opinions A or B, and it is hard to stay in the not sure AB state. The model is then unstable, because agents switch between A and B very easily. At the extreme value of $\gamma = 0.99$ the behavior changes and the not sure state becomes absorbing, since a sharp opinion requires one weight close to one and the opposite close to zero, so any interaction can push the weight difference below the $\gamma$ threshold.

Increasing $\gamma$ raises the overall F1 and lowers the change F1 on every topic. With a small $\mu$, a larger $\gamma$ makes opinion change harder, so more agents keep their initial opinion. This is consistent with the data, where most opinions do not change, and it is why the trajectory JSD improves together with the overall F1, as the model better reproduces the empirical distribution of opinion transitions. Job guarantee shows the limit of this effect, since more than forty percent of its agents change their opinion. There the trajectory JSD improves only up to a point and then deteriorates, while the overall F1 keeps growing, because a model that rarely changes opinion no longer matches the empirical transitions.

\begin{figure}[ht]
    \centering
    \subfloat[\label{fig:sens_fssocsec}]
    {\includegraphics[scale=0.34]{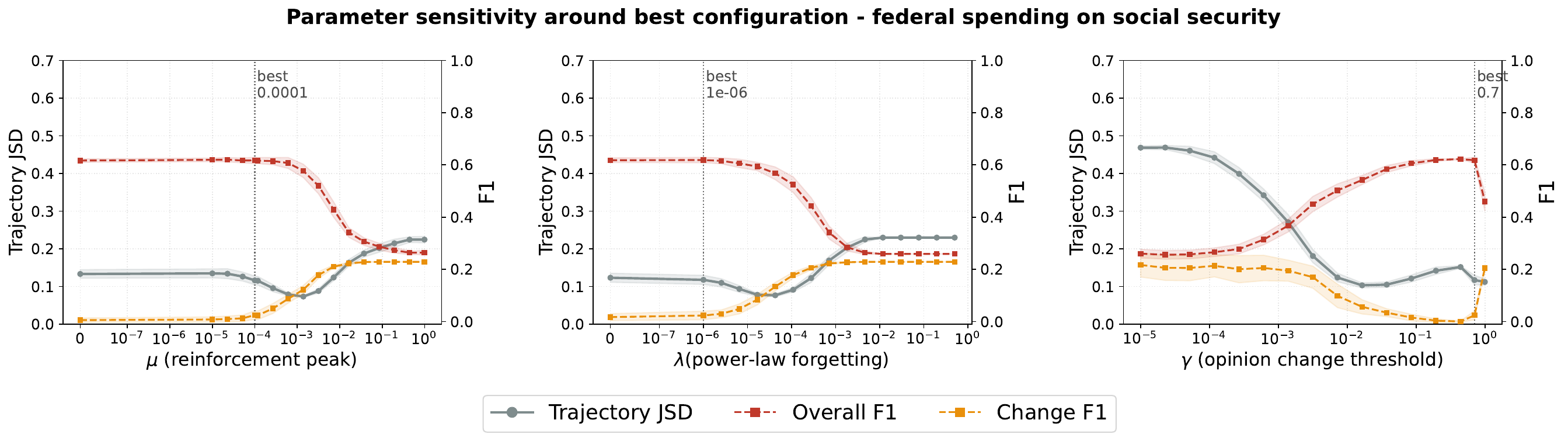}}
    \vspace{1em}
    \subfloat[\label{fig:sens_jobguar}]
    {\includegraphics[scale=0.34]{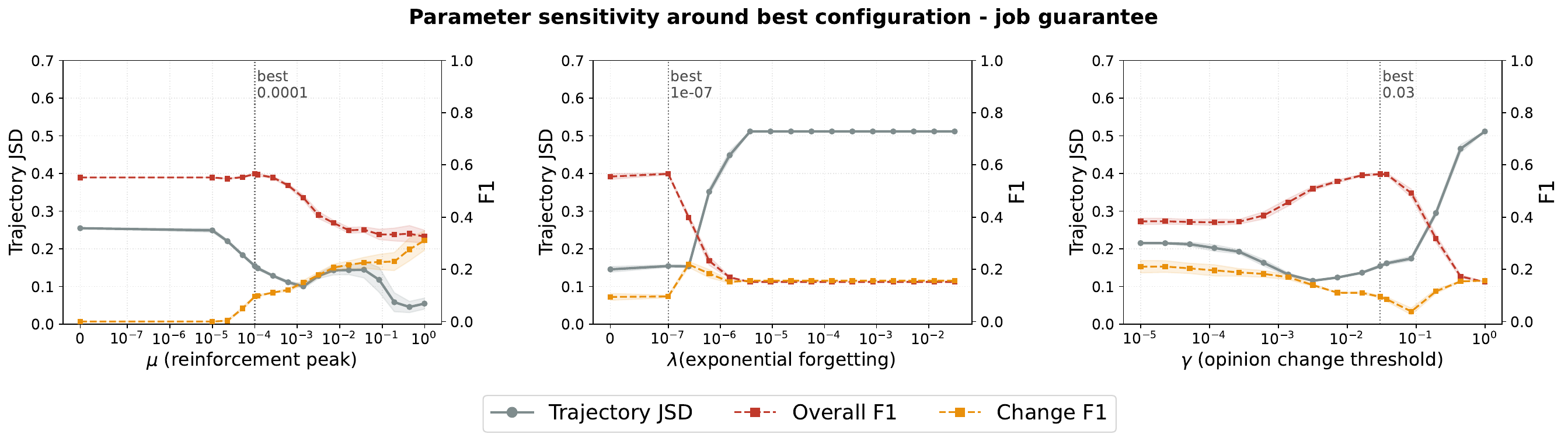}}
    \caption{Parameter sensitivity around the best configuration for two representative topics: (a)~federal spending on social security (power-law forgetting) and (b)~job guarantee (exponential forgetting). In each subfigure, one parameter ($\mu$, $\lambda$, or $\gamma$ )is varied while the others are fixed at their best values. The vertical dashed line marks the best value of the varied parameter. Each panel reports the trajectory JSD on the left axis (lower is better) and the overall and change F1 scores on the right axis (higher is better).}
\end{figure}

\begin{table}[t]
\centering
\footnotesize
\caption{Effect size (rank-biserial correlation, RBC) of CoDiNG versus each
baseline, per topic and metric. Positive values indicate a CoDiNG advantage,
negative a baseline advantage; bold marks a significant advantage ($p_{\text{BH}} < 0.05$, one-sided Wilcoxon signed-rank test, Benjamini- Hochberg-corrected within each metric family).}
\label{tab:stat_tests}
 
\noindent
\begin{minipage}[t]{0.49\textwidth}\centering
\begin{tabular}{l|cc|cc|c}
\multicolumn{6}{c}{\textbf{Euthanasia}}\\
\midrule
& \multicolumn{2}{c|}{Distrib. JSD} & \multicolumn{2}{c|}{Traject. JSD} & Overall \\
& S5 & S6 & S4$\to$S5 & S4$\to$S6 & F1\\
\midrule
NG & \bfw{+1.00} & \bfw{+1.00} & \bfw{+1.00} & \bfw{+1.00} & \bfw{+1.00}\\
SJBO & \bfw{+1.00} & -0.97 & \bfw{+1.00} & \bfw{+1.00} & -0.82\\
Keep & \bfw{+1.00} & -1.00 & \bfw{+1.00} & \bfw{+1.00} & -1.00\\
Majority & \bfw{+1.00} & \bfw{+1.00} & \bfw{+1.00} & \bfw{+1.00} & \bfw{+1.00}\\
Rand-U & \bfw{+0.94} & \bfw{+0.33} & \bfw{+1.00} & \bfw{+1.00} & \bfw{+1.00}\\
Rand-E & \bfw{+0.94} & -0.72 & \bfw{+1.00} & \bfw{+0.99} & \bfw{+1.00}\\
\end{tabular}
\end{minipage}\hfill
\begin{minipage}[t]{0.49\textwidth}\centering
\begin{tabular}{l|cc|cc|c}
\multicolumn{6}{c}{\textbf{Federal spending on social security}}\\
\midrule
& \multicolumn{2}{c|}{Distrib. JSD} & \multicolumn{2}{c|}{Traject. JSD} & Overall \\
& S5 & S6 & S4$\to$S5 & S4$\to$S6 & F1\\
\midrule
NG & \bfw{+1.00} & \bfw{+1.00} & \bfw{+1.00} & \bfw{+1.00} & \bfw{+1.00}\\
SJBO & \bfw{+0.38} & -0.93 & \bfw{+0.44} & \bfw{+0.64} & \bfw{+0.55}\\
Keep & \bfw{+0.75} & -0.28 & \bfw{+1.00} & \bfw{+1.00} & -0.72\\
Majority & \bfw{+1.00} & \bfw{+1.00} & \bfw{+1.00} & \bfw{+1.00} & \bfw{+1.00}\\
Rand-U & \bfw{+1.00} & \bfw{+1.00} & \bfw{+1.00} & \bfw{+0.94} & \bfw{+1.00}\\
Rand-E & \bfw{+0.92} & -0.24 & -0.90 & -0.89 & \bfw{+1.00}\\
\end{tabular}
\end{minipage}
\vspace{1.5ex}
 
\noindent
\begin{minipage}[t]{0.49\textwidth}\centering
\begin{tabular}{l|cc|cc|c}
\multicolumn{6}{c}{\textbf{Federal spending on welfare}}\\
\midrule
& \multicolumn{2}{c|}{Distrib. JSD} & \multicolumn{2}{c|}{Traject. JSD} & Overall \\
& S5 & S6 & S4$\to$S5 & S4$\to$S6 & F1\\
\midrule
NG & \bfw{+1.00} & \bfw{+1.00} & \bfw{+1.00} & \bfw{+1.00} & \bfw{+1.00}\\
SJBO & \bfw{+1.00} & \bfw{+1.00} & \bfw{+0.91} & \bfw{+0.97} & \bfw{+0.33}\\
Keep & \bfw{+1.00} & \bfw{+1.00} & \bfw{+1.00} & \bfw{+1.00} & -0.73\\
Majority & \bfw{+1.00} & \bfw{+1.00} & \bfw{+1.00} & \bfw{+1.00} & \bfw{+1.00}\\
Rand-U & \bfw{+1.00} & \bfw{+1.00} & \bfw{+0.97} & \bfw{+0.99} & \bfw{+1.00}\\
Rand-E & \bfw{+0.84} & \bfw{+0.65} & -0.75 & -0.30 & \bfw{+1.00}\\
\end{tabular}
\end{minipage}\hfill
\begin{minipage}[t]{0.49\textwidth}\centering
\begin{tabular}{l|cc|cc|c}
\multicolumn{6}{c}{\textbf{Job guarantee}}\\
\midrule
& \multicolumn{2}{c|}{Distrib. JSD} & \multicolumn{2}{c|}{Traject. JSD} & Overall \\
& S5 & S6 & S4$\to$S5 & S4$\to$S6 & F1\\
\midrule
NG & \bfw{+1.00} & \bfw{+1.00} & -0.11 & \bfw{+1.00} & \bfw{+1.00}\\
SJBO & -0.85 & -0.21 & -0.94 & \bfw{+0.67} & \bfw{+0.99}\\
Keep & \bfw{+1.00} & \bfw{+1.00} & \bfw{+1.00} & \bfw{+1.00} & \bfw{+1.00}\\
Majority & \bfw{+1.00} & \bfw{+1.00} & \bfw{+1.00} & \bfw{+1.00} & \bfw{+1.00}\\
Rand-U & \bfw{+0.99} & \bfw{+1.00} & -1.00 & -0.98 & \bfw{+1.00}\\
Rand-E & \bfw{+0.94} & \bfw{+0.99} & -1.00 & -1.00 & \bfw{+1.00}\\
\end{tabular}
\end{minipage}
\vspace{1.5ex}
 
\noindent
\begin{minipage}[t]{0.49\textwidth}\centering
\begin{tabular}{l|cc|cc|c}
\multicolumn{6}{c}{\textbf{Marijuana}}\\
\midrule
& \multicolumn{2}{c|}{Distrib. JSD} & \multicolumn{2}{c|}{Traject. JSD} & Overall \\
& S5 & S6 & S4$\to$S5 & S4$\to$S6 & F1\\
\midrule
NG & \bfw{+1.00} & \bfw{+1.00} & \bfw{+1.00} & \bfw{+1.00} & \bfw{+1.00}\\
SJBO & -1.00 & -1.00 & \bfw{+0.68} & \bfw{+0.80} & -1.00\\
Keep & -1.00 & -1.00 & \bfw{+1.00} & \bfw{+1.00} & -1.00\\
Majority & \bfw{+1.00} & \bfw{+1.00} & \bfw{+1.00} & \bfw{+1.00} & \bfw{+1.00}\\
Rand-U & \bfw{+0.82} & -0.88 & \bfw{+1.00} & +0.18 & \bfw{+1.00}\\
Rand-E & -0.32 & -0.85 & \bfw{+1.00} & \bfw{+0.31} & \bfw{+1.00}\\
\end{tabular}
\end{minipage}\hfill
\begin{minipage}[t]{0.49\textwidth}\centering
\begin{tabular}{l|cc|cc|c}
\multicolumn{6}{c}{\textbf{Equal rights}}\\
\midrule
& \multicolumn{2}{c|}{Distrib. JSD} & \multicolumn{2}{c|}{Traject. JSD} & Overall \\
& S5 & S6 & S4$\to$S5 & S4$\to$S6 & F1\\
\midrule
NG & \bfw{+1.00} & \bfw{+1.00} & \bfw{+1.00} & \bfw{+1.00} & \bfw{+1.00}\\
SJBO & \bfw{+0.95} & \bfw{+0.96} & -0.56 & +0.02 & -0.18\\
Keep & \bfw{+1.00} & \bfw{+1.00} & \bfw{+1.00} & \bfw{+1.00} & -0.99\\
Majority & \bfw{+1.00} & \bfw{+1.00} & \bfw{+1.00} & \bfw{+1.00} & \bfw{+1.00}\\
Rand-U & \bfw{+1.00} & \bfw{+0.98} & -0.64 & \bfw{+0.30} & \bfw{+1.00}\\
Rand-E & \bfw{+0.95} & \bfw{+0.83} & -0.87 & +0.18 & \bfw{+1.00}\\
\end{tabular}
\end{minipage}
\vspace{1.5ex}
 
\vspace{0.5ex}
\centerline{\begin{tabular}{l|cc|cc|c}
\toprule
\multicolumn{6}{c}{\textbf{CoDiNG significant wins (out of 6 topics)}}\\
\midrule
Baseline & \multicolumn{2}{c|}{Distrib. JSD} & \multicolumn{2}{c|}{Traject. JSD} & Overall \\
& S5 & S6 & S4$\to$S5 & S4$\to$S6 & F1\\
\midrule
\textbf{NG} & \textbf{6/6} & \textbf{6/6} & \textbf{5/6} & \textbf{6/6} & \textbf{6/6}\\
\textbf{SJBO} & \textbf{4/6} & \textbf{2/6} & \textbf{4/6} & \textbf{5/6} & \textbf{3/6}\\
\textbf{Keep} & \textbf{5/6} & \textbf{3/6} & \textbf{6/6} & \textbf{6/6} & \textbf{1/6}\\
\textbf{Majority} & \textbf{6/6} & \textbf{6/6} & \textbf{6/6} & \textbf{6/6} & \textbf{6/6}\\
\textbf{Random (uniform)} & \textbf{6/6} & \textbf{5/6} & \textbf{4/6} & \textbf{4/6} & \textbf{6/6}\\
\textbf{Random (empirical)} & \textbf{5/6} & \textbf{3/6} & \textbf{2/6} & \textbf{2/6} & \textbf{6/6}\\
\bottomrule
\end{tabular}}
\end{table}

\section{Conclusions}
\label{sec:conclusions}
We proposed CoDiNG as an extended Naming Game model where each agent holds a continuous latent state driven by reinforcement and forgetting mechanisms. We evaluated this model on real interaction and survey data from the NetSense study. In our dataset opinions change at the individual level while the aggregate distribution remains almost stationary. Therefore a model that reproduces no dynamics at all can still achieve a high score.

Evaluating opinion models based on a single metric is therefore misleading because model rankings reverse depending on whether overall accuracy or change detection is measured. This tension is present in the baselines themselves since the Keep baseline maximizes overall accuracy while detecting no change at all and the empirical random baseline does the exact opposite. Only hybrid models possess the parametric flexibility required to balance overall accuracy with change dynamics. However a choice regarding which objective to maximize must be made since no model excels at both simultaneously. CoDiNG consistently outperforms or remains competitive with the SJBO model. This demonstrates that reinforcement and forgetting mechanisms derived from cognitive science can contribute to the explanation of opinion dynamics.

Our findings align with recent work reporting that opinion models calibrated on empirical data learn to freeze rather than reproduce true dynamics \cite{moor2026testing}. That previous work relies on cross sectional surveys and validates models solely against aggregate distributions while our data tracks the same individuals over time. Opinion dynamics models remain valuable for studying generative mechanisms such as consensus and polarization but on real world data they often fail to outperform trivial baselines. The field of opinion dynamics would greatly benefit from devoting more attention to empirical validation.

A primary limitation of our work is that the sensitivity analysis is local and therefore the observed behavior of CoDiNG cannot be broadly generalized. A natural step for future research is to study CoDiNG on synthetic data to characterize its general properties. These include the conditions under which it reaches consensus or polarization and whether the sharp change in behavior around the opinion change threshold corresponds to a phase transition.

Another promising avenue for extending the CoDiNG model involves moving beyond binary options. We can instead adapt the model for multidimensional environments where opinions cannot be easily mapped as direct opposites. This could potentially include the exploration of emotional associations linked to specific topics.

\section{Data Availability Statement}
The NetSense dataset is available upon request from Prof. Omar Lizardo. The code and the data samples are available online as a Code Ocean capsule at \href{https://doi.org/10.24433/CO.8066140.v1}{https://doi.org/10.24433/CO.8066140.v1}. 

\section*{Acknowledgments}
The Polish National Science Centre supported this work under grant no. 2021/41/B/HS6/02798. This work was also partially funded by the Polish Ministry of Education and Science within the “International Projects Co-Funded” program and the European Union under the Horizon Europe, grant no. 101086321 (OMINO). The DARPA-INCAS Program is under Agreement No HR001121C0165 and the NSF Grant No. BSE-2214216 partially supported BKS. However, the views and opinions expressed are those of the author(s) only and do not necessarily reflect those of the European Union or the European Research Executive Agency. Neither the European Union nor the European Research Executive Agency can be held responsible for them. To create the results, we used the resources provided by the Wroclaw Centre for Networking and Supercomputing (http://wcss.pl).

\bibliographystyle{IEEEtran}
\bibliography{references} 

@book{Mace2019-gl,
  title     = "Organization and Structure of Autobiographical Memory",
  author    = "Mace, John",
  publisher = "Oxford University Press",
  year      =  2019
}

@book{De_Saussure1964-pz,
  title     = {Course in General Linguistics},
  author    = {de Saussure, Ferdinand},
  editor    = {{F} and Bally, C and Sechehaye, A},
  publisher = {Open Court},
  address   = {Chicago},
  year      =  {1964}
}

@article{bahulkar2017coevolution,
  title={Coevolution of a multilayer node-aligned network whose layers represent different social relations},
  author={Bahulkar, Ashwin and Szymanski, Boleslaw K and Chan, Kevin and Lizardo, Omar},
  journal={Computational social networks},
  volume={4},
  pages={1--22},
  year={2017},
  publisher={Springer}
}

@book{atkin1977combinatorial,
  title={Combinatorial Connectivities in Social Systems: an application of simplicial complex structures to the study of large organizations},
  author={Atkin, Ronald H},
  year={1977},
  publisher={Springer}
}

@article{dunbar2007evolution,
  title={Evolution in the social brain},
  author={Dunbar, Robin IM and Shultz, Susanne},
  journal={Science},
  volume={317},
  number={5843},
  pages={1344--1347},
  year={2007},
  publisher={American Association for the Advancement of Science}
}

@book{tarde1899,
  title={Social laws: An outline of sociology},
  author={De Tarde, Gabriel},
  year={1899},
  publisher={Macmillan}
}

@article{granovetter1978threshold,
  title={Threshold models of collective behavior},
  author={Granovetter, Mark},
  journal={American Journal of Sociology},
  volume={83},
  number={6},
  pages={1420--1443},
  year={1978}
}

@article{gleeson2013binary,
  title={Binary-state dynamics on complex networks: Pair approximation and beyond},
  author={Gleeson, James P},
  journal={Physical Review X},
  volume={3},
  number={2},
  pages={021004},
  year={2013},
  publisher={APS}
}

@article{pierce2001global,
  title={Global and specific relational models in the experience of social interactions.},
  author={Pierce, Tamarha and Lydon, John E},
  journal={Journal of personality and social psychology},
  volume={80},
  number={4},
  pages={613},
  year={2001},
  publisher={American Psychological Association}
}

@article{karsai2011small,
  title={Small but slow world: How network topology and burstiness slow down spreading},
  author={Karsai, M{\'a}rton and Kivel{\"a}, Mikko and Pan, Raj Kumar and Kaski, Kimmo and Kert{\'e}sz, J{\'a}nos and Barab{\'a}si, A-L and Saram{\"a}ki, Jari},
  journal={Physical Review E},
  volume={83},
  number={2},
  pages={025102},
  year={2011},
  publisher={APS}
}

@article{watts1998collective,
  title={Collective dynamics of ‘small-world’networks},
  author={Watts, Duncan J and Strogatz, Steven H},
  journal={Nature},
  volume={393},
  number={6684},
  pages={440--442},
  year={1998},
  publisher={Nature Publishing Group}
}

@article{albert2002statistical,
  title={Statistical mechanics of complex networks},
  author={Albert, R{\'e}ka and Barab{\'a}si, Albert-L{\'a}szl{\'o}},
  journal={Reviews of modern physics},
  volume={74},
  number={1},
  pages={47},
  year={2002},
  publisher={APS}
}

@article{guilbeault2018complex,
  title={Complex contagions: A decade in review},
  author={Guilbeault, Douglas and Becker, Joshua and Centola, Damon},
  journal={Complex spreading phenomena in social systems: Influence and contagion in real-world social networks},
  pages={3--25},
  year={2018},
  publisher={Springer}
}

@article{friedkin1997social,
  title={Social positions in influence networks},
  author={Friedkin, Noah E and Johnsen, Eugene C},
  journal={Social networks},
  volume={19},
  number={3},
  pages={209--222},
  year={1997},
  publisher={Elsevier}
}

@article{moran1950notes,
  title={Notes on continuous stochastic phenomena},
  author={Moran, Patrick AP},
  journal={Biometrika},
  volume={37},
  number={1/2},
  pages={17--23},
  year={1950},
  publisher={JSTOR}
}

@article{bass1994bass,
  title={Why the Bass model fits without decision variables},
  author={Bass, Frank M and Krishnan, Trichy V and Jain, Dipak C},
  journal={Marketing science},
  volume={13},
  number={3},
  pages={203--223},
  year={1994},
  publisher={INFORMS}
}

@book{carrington2005models,
  title={Models and methods in social network analysis},
  author={Carrington, Peter J and Scott, John and Wasserman, Stanley},
  volume={28},
  year={2005},
  publisher={Cambridge university press}
}

@article{centola2015social,
  title={The social origins of networks and diffusion},
  author={Centola, Damon},
  journal={American Journal of Sociology},
  volume={120},
  number={5},
  pages={1295--1338},
  year={2015},
  publisher={University of Chicago Press Chicago, IL}
}

@book{friedkin2011social,
  title={Social influence network theory: A sociological examination of small group dynamics},
  author={Friedkin, Noah E and Johnsen, Eugene C},
  volume={33},
  year={2011},
  publisher={Cambridge University Press}
}

@inproceedings{zbieg2012studying,
  title={Studying diffusion of viral content at dyadic level},
  author={Zbieg, Anita and Zak, Blazej and Jankowski, Jaroslaw and Michalski, Radoslaw and Ciuberek, Sylwia},
  booktitle={2012 IEEE/ACM International Conference on Advances in Social Networks Analysis and Mining},
  pages={1259--1265},
  year={2012},
  organization={IEEE}
}

@article{friedkin1990social,
  title={Social influence and opinions},
  author={Friedkin, Noah E and Johnsen, Eugene C},
  journal={Journal of mathematical sociology},
  volume={15},
  number={3-4},
  pages={193--206},
  year={1990},
  publisher={Taylor \& Francis}
}

@article{rogers1994diffusion,
  title={Diffusion of innovations, 1995},
  author={Rogers, Everett M},
  journal={New York: Free Pr},
  year={1994}
}

@book{huttenlocher2009neural,
  title={Neural Plasticity: the Effects of Environment on the Development of the Cerebral Cortex},
  author={Huttenlocher, Peter R},
  year={2009},
  publisher={Harvard University Press}
}

@Article{baddeley2004psychology,
  Title                    = {The psychology of memory},
  Author                   = {Baddeley, Alan D},
  Journal                  = {The Essential Handbook of Memory Disorders for Clinicians},
  Year                     = {2004},
  Pages                    = {1--13},
  Publisher                = {John Wiley \& Sons, Ltd}
}

@Article{anderson1998integrated,
  Title                    = {An integrated theory of list memory},
  Author                   = {Anderson, John R and Bothell, Dan and Lebiere, Christian and Matessa, Michael},
  Journal                  = {Journal of Memory and Language},
  Year                     = {1998},
  Number                   = {4},
  Pages                    = {341--380},
  Volume                   = {38},
  Publisher                = {Elsevier}
}

@article{anderson1997act,
  title={ACT-R: A theory of higher level cognition and its relation to visual attention},
  author={Anderson, John R and Matessa, Michael and Lebiere, Christian},
  journal={Human--Computer Interaction},
  volume={12},
  number={4},
  pages={439--462},
  year={1997},
  publisher={Taylor \& Francis}
}

@article{holyst2024protect,
  title={Protect our environment from information overload},
  author={Ho{\l}yst, Janusz A and Mayr, Philipp and Thelwall, Michael and Frommholz, Ingo and Havlin, Shlomo and Sela, Alon and Kenett, Yoed N and Helic, Denis and Rehar, Aljo{\v{s}}a and Ma{\v{c}}ek, Sebastijan R and others},
  journal={Nature Human Behaviour},
  pages={1--2},
  year={2024},
  publisher={Nature Publishing Group UK London}
}

@article{mills1940situated,
  title={Situated actions and vocabularies of motive},
  author={Mills, C Wright},
  journal={American sociological review},
  volume={5},
  number={6},
  pages={904--913},
  year={1940},
  publisher={JSTOR}
}

@article{michalski2021social,
  title={Social networks through the prism of cognition},
  author={Michalski, Rados{\l}aw and Szymanski, Boleslaw K and Kazienko, Przemys{\l}aw and Lebiere, Christian and Lizardo, Omar and Kulisiewicz, Marcin},
  journal={Complexity},
  volume={2021},
  pages={1--13},
  year={2021},
  publisher={Hindawi Limited}
}

@article{xie2011social,
  title={Social consensus through the influence of committed minorities},
  author={Xie, Jierui and Sreenivasan, Sameet and Korniss, Gyorgy and Zhang, Weituo and Lim, Chjan and Szymanski, Boleslaw K},
  journal={Physical Review E},
  volume={84},
  number={1},
  pages={011130},
  year={2011},
  publisher={APS}
}

@article{xie2012evolution,
  title={Evolution of opinions on social networks in the presence of competing committed groups},
  author={Xie, Jierui and Emenheiser, Jeffrey and Kirby, Matthew and Sreenivasan, Sameet and Szymanski, Boleslaw K and Korniss, Gyorgy},
  journal={PLoS One},
  volume={7},
  number={3},
  pages={e33215},
  year={2012},
  publisher={Public Library of Science San Francisco, USA}
}

@article{striegel2013lessons,
  title={Lessons learned from the netsense smartphone study},
  author={Striegel, Aaron and Liu, Shu and Meng, Lei and Poellabauer, Christian and Hachen, David and Lizardo, Omar},
  journal={ACM SIGCOMM Computer Communication Review},
  volume={43},
  number={4},
  pages={51--56},
  year={2013},
  publisher={ACM New York, NY, USA}
}

@article{degroot1974reaching,
  title={Reaching a consensus},
  author={DeGroot, Morris H},
  journal={Journal of the American Statistical association},
  volume={69},
  number={345},
  pages={118--121},
  year={1974},
  publisher={Taylor \& Francis}
}

@article{deffuant2000mixing,
  title={Mixing beliefs among interacting agents},
  author={Deffuant, Guillaume and Neau, David and Amblard, Frederic and Weisbuch, G{\'e}rard},
  journal={Advances in Complex Systems},
  volume={3},
  number={01n04},
  pages={87--98},
  year={2000},
  publisher={World Scientific}
}

@article{rainer2002opinion,
  title={Opinion dynamics and bounded confidence: models, analysis and simulation},
  author={Rainer, Hegselmann and Krause, Ulrich},
  journal={Journal of Artificial Societies and Social Simulation},
  volume = {5},
  number = {3},
  pages={1--30},
  year={2002}
}

@article{martins2008continuous,
  title={Continuous opinions and discrete actions in opinion dynamics problems},
  author={Martins, Andr{\'e} CR},
  journal={International Journal of Modern Physics C},
  volume={19},
  number={04},
  pages={617--624},
  year={2008},
  publisher={World Scientific}
}

@article{zhan2021bounded,
  title={Bounded confidence evolution of opinions and actions in social networks},
  author={Zhan, Min and Kou, Gang and Dong, Yucheng and Chiclana, Francisco and Herrera-Viedma, Enrique},
  journal={IEEE Transactions on Cybernetics},
  volume={52},
  number={7},
  pages={7017--7028},
  year={2021},
  publisher={IEEE}
}

@article{fan2016opinion,
  title={Opinion evolution influenced by informed agents},
  author={Fan, Kangqi and Pedrycz, Witold},
  journal={Physica A: Statistical Mechanics and Its Applications},
  volume={462},
  pages={431--441},
  year={2016},
  publisher={Elsevier}
}

@book{zaller1992nature,
  title={The nature and origins of mass opinion},
  author={Zaller, John},
  year={1992},
  publisher={Cambridge university press}
}

@article{baronchelli2016gentle,
  title={A gentle introduction to the minimal naming game},
  author={Baronchelli, Andrea},
  journal={Belgian Journal of Linguistics},
  volume={30},
  number={1},
  pages={171--192},
  year={2016},
  publisher={John Benjamins Publishing Company Amsterdam/Philadelphia}
}

@article{davis1963structural,
  title={Structural balance, mechanical solidarity, and interpersonal relations},
  author={Davis, James A},
  journal={American journal of sociology},
  volume={68},
  number={4},
  pages={444--462},
  year={1963},
  publisher={University of Chicago Press}
}

@article{valente1996social,
  title={Social network thresholds in the diffusion of innovations},
  author={Valente, Thomas W},
  journal={Social networks},
  volume={18},
  number={1},
  pages={69--89},
  year={1996},
  publisher={Elsevier}
}

@article{fortunato2013statistical,
  title={Statistical mechanics and social sciences},
  author={Fortunato, Santo and Macy, Michael and Redner, Sidney},
  journal={arXiv preprint arXiv:1304.1171},
  year={2013}
}

@book{neisser2000memory,
  title={Memory observed: Remembering in natural contexts},
  author={Neisser, Ulric and Hyman, Ira},
  year={2000},
  publisher={Macmillan}
}

@article{squire2004memory,
  title={Memory systems of the brain: a brief history and current perspective},
  author={Squire, Larry R},
  journal={Neurobiology of learning and memory},
  volume={82},
  number={3},
  pages={171--177},
  year={2004},
  publisher={Elsevier}
}

@article{goldenberg2001talk,
  title={Talk of the network: A complex systems look at the underlying process of word-of-mouth},
  author={Goldenberg, Jacob and Libai, Barak and Muller, Eitan},
  journal={Marketing letters},
  volume={12},
  pages={211--223},
  year={2001},
  publisher={Springer}
}

@article{steels1995self,
  title={A self-organizing spatial vocabulary},
  author={Steels, Luc},
  journal={Artificial life},
  volume={2},
  number={3},
  pages={319--332},
  year={1995}
}

@article{clifford1973model,
  title={A model for spatial conflict},
  author={Clifford, Peter and Sudbury, Aidan},
  journal={Biometrika},
  volume={60},
  number={3},
  pages={581--588},
  year={1973},
  publisher={Oxford University Press}
}

@article{galam1999application,
  title={Application of statistical physics to politics},
  author={Galam, Serge},
  journal={Physica A: Statistical mechanics and its applications},
  volume={274},
  number={1-2},
  pages={132--139},
  year={1999},
  publisher={Elsevier}
}

@article{baronchelli2006sharp,
  title={Sharp transition towards shared vocabularies in multi-agent systems},
  author={Baronchelli, Andrea and Felici, Maddalena and Loreto, Vittorio and Caglioti, Emanuele and Steels, Luc},
  journal={Journal of Statistical Mechanics: Theory and Experiment},
  volume={2006},
  number={06},
  pages={P06014},
  year={2006},
  publisher={IOP Publishing}
}

@article{sznajd2000opinion,
  title={Opinion evolution in closed community},
  author={Sznajd-Weron, Katarzyna and Sznajd, Jozef},
  journal={International Journal of Modern Physics C},
  volume={11},
  number={06},
  pages={1157--1165},
  year={2000},
  publisher={World Scientific}
}

@article{ising1925beitrag, 
  title={Beitrag zur Theorie des Ferromagnetismus}, 
  author={Ising, Ernst}, 
  journal={Zeitschrift f{\"u}r Physik}, 
  volume={31}, 
  number={1}, 
  pages={253--258}, 
  year={1925}, 
  publisher={Springer} 
}

@article{stauffer2000generalization,
  title={Generalization to square lattice of Sznajd sociophysics model},
  author={Stauffer, Dietrich and Sousa, Alexandra O and De Oliveira, S Moss},
  journal={International Journal of Modern Physics C},
  volume={11},
  number={06},
  pages={1239--1245},
  year={2000},
  publisher={World Scientific}
}

@article{bernardes2001damage,
  title={Damage spreading, coarsening dynamics and distribution of political votes in Sznajd model on square lattice},
  author={Bernardes, AT and Costa, UMS and Araujo, AD and Stauffer, D},
  journal={International Journal of Modern Physics C},
  volume={12},
  number={02},
  pages={159--167},
  year={2001},
  publisher={World Scientific}
}

@article{bernardes2002election,
  title={Election results and the Sznajd model on Barabasi network},
  author={Bernardes, Americo T and Stauffer, Dietrich and Kert{\'e}sz, Janos},
  journal={The European Physical Journal B-Condensed Matter and Complex Systems},
  volume={25},
  pages={123--127},
  year={2002},
  publisher={Springer}
}

@article{elgazzar2001application,
  title={Application of the Sznajd sociophysics model to small-world networks},
  author={Elgazzar, AS},
  journal={International Journal of Modern Physics C},
  volume={12},
  number={10},
  pages={1537--1544},
  year={2001},
  publisher={World Scientific}
}

@article{ru2008opinion,
  title={Opinion dynamics on complex networks with communities},
  author={Ru, Wang and Li-Ping, Chi},
  journal={Chinese Physics Letters},
  volume={25},
  number={4},
  pages={1502},
  year={2008},
  publisher={IOP Publishing}
}

@article{moor2026testing,
  title={Testing the validity of multiple opinion dynamics models},
  author={Moor-Smith, Samuel and Carpentras, Dino},
  journal={arXiv preprint arXiv:2602.00876},
  year={2026}
}

@article{fan2015emergence,
  title={Emergence and spread of extremist opinions},
  author={Fan, Kangqi and Pedrycz, Witold},
  journal={Physica A: Statistical Mechanics and its Applications},
  volume={436},
  pages={87--97},
  year={2015},
  publisher={Elsevier}
}


 



\vfill

\appendix
\renewcommand{\thefigure}{\thesection.\arabic{figure}}
\renewcommand{\thetable}{\thesection.\arabic{table}}
\setcounter{figure}{0}
\setcounter{table}{0}
\counterwithin{figure}{section}
\counterwithin{table}{section}

\section{Parameter Space}
\label{app:parameter-space}
Table~\ref{tab:coding-param-grid} lists the full parameter search space explored during the calibration of CoDiNG described in Section~\ref{subsec:setting} of the main article. The random search drew $n_{\mathrm{iter}} = 200$ combinations from this grid for each topic. Table~\ref{tab:sjbo-param-grid} reports the corresponding search space for the SJBO model, which we search with the same protocol to ensure a fair comparison.

\begin{table}[!ht]
\centering
\caption{Parameter search space explored by random search ($n_{\mathrm{iter}}=200$).}
\label{tab:coding-param-grid}
\begin{tabular}{lcl}
\toprule
\textbf{Parameter} & \textbf{Symbol} & \textbf{Tested values} \\
\midrule
Forgetting kernel & -- & $\{\mathrm{pow},\ \mathrm{exp}\}$ \\
Reinforcement peak & $\mu$ & $\{10^{-4},\ 5{\cdot}10^{-4},\ 10^{-3}\} \cup \{0.005,\ 0.01,\ \dots,\ 0.045\} \cup \{0.05,\ 0.1,\ \dots,\ 0.6\}$ \\
Neutrality threshold & $\gamma$ & $\{10^{-3},\ 5{\cdot}10^{-3}\} \cup \{0.01,\ 0.02,\ \dots,\ 0.09\} \cup \{0.1,\ 0.15,\ \dots,\ 0.7\}$ \\
Forgetting parameter (exp) & $\lambda$ & $\{10^e \mid e \in \{-7,\ -6.5,\ \dots,\ -3.5\} \cup \{-3,\ -2.75,\ \dots,\ -1.75\}\}$ \\
Forgetting parameter (pow) & $\lambda$ & $\{10^e \mid e \in \{-6,\ -5.5,\ \dots,\ -1.5\}\} \cup \{0.05,\ 0.1,\ \dots,\ 0.5\}$ \\
\bottomrule
\end{tabular}
\end{table}

\begin{table}[!ht]
\centering
\caption{Parameter search space explored by random search for the SJBO model ($n_{\mathrm{iter}}=200$).}
\label{tab:sjbo-param-grid}
\begin{tabular}{lcl}
\toprule
\textbf{Parameter} & \textbf{Symbol} & \textbf{Tested values} \\
\midrule
Assimilation threshold & $\epsilon$ & $\{0.6, 0.8, 0.9\} \cup \{1.0,\ 1.05,\ \dots,\ 1.9\}$ \\
Assimilation rate & $a$ & $\{0.01,\ 0.05\} \cup \{0.1,\ 0.15,\ \dots,\ 0.5\}$ \\
Rejection rate & $r$ & $\{0.1,\ 0.15,\ \dots,\ 1.0\}$ \\
Hesitation / silence threshold & $h$ & $\{0.1,\ 0.15,\ \dots,\ 0.8\}$ \\
Decay threshold & $\rho$ & $\{0.1,\ 0.15,\ \dots,\ 0.8\}$ \\
Decay multiplier & $\lambda$ & $\{0.4,\ 0.45,\ \dots,\ 0.95\}$ \\
Rejection threshold & $\tau$ & $\min(2.0, \epsilon + \Delta\tau) \text{ where } \Delta\tau \in \{0.2,\ 0.3,\ \dots,\ 0.8\}$ \\
\bottomrule
\end{tabular}
\end{table}

\section{The Best Parameters}
\label{app:best-parameters}
Table~\ref{tab:best_params} reports the parameter configuration of CoDiNG model selected for each topic and each sample weight assigned to opinion-change cases ($w_1$ to $w_{80}$), obtained from the search described in Section~\ref{subsec:setting} of the main article. These are the configurations underlying the Pareto fronts in Section~VI-B (Fig.~\ref{fig:pareto} of the main article), which illustrate the trade-off between overall F1 and change F1 as the sample weight increases. Similarly, Table~\ref{tab:best_params_sjbo} presents the best parameters for the SJBO model.

\begin{table}[!ht]
\caption{The best parameters - CoDiNG}
\label{tab:best_params}
\setlength{\extrarowheight}{2pt}
\centering
\begin{tabular}{|c|c|c|c|c|c|}
\hline
\multirow{2}{*}{\textbf{Sample weights}} & \multirow{2}{*}{\textbf{Topics}} & \multicolumn{4}{c|}{\textbf{The best parameters}} \\
\cline{3-6}
 &  & $\mu$ & $\lambda$ & $\gamma$ & forget. \\
\hline
\multirow{6}{*}{1:1 (w1)} & Euthanasia & 0.0001 & $10^{-6}$ & 0.7 & pow \\
\cline{2-6}
 & Fed. spend. on social security & 0.0001 & $10^{-6}$ & 0.7 & pow \\
\cline{2-6}
 & Fed. spend. on welfare & 0.0001 & $10^{-6}$ & 0.7 & pow \\
\cline{2-6}
 & Job guarantee & 0.0001 & $10^{-7}$ & 0.03 & exp \\
\cline{2-6}
 & Marijuana & 0.0001 & $10^{-6}$ & 0.7 & pow \\
\cline{2-6}
 & Equal rights & 0.0001 & $10^{-6}$ & 0.7 & pow \\
\hline\hline
\multirow{6}{*}{1:3 (w3)} & Euthanasia & 0.025 & $3.2\times10^{-5}$ & 0.005 & pow \\
\cline{2-6}
 & Fed. spend. on social security & 0.025 & $3.2\times10^{-5}$ & 0.07 & pow \\
\cline{2-6}
 & Fed. spend. on welfare & 0.025 & $3.2\times10^{-5}$ & 0.005 & pow \\
\cline{2-6}
 & Job guarantee & 0.6 & 0.5 & 0.6 & pow \\
\cline{2-6}
 & Marijuana & 0.005 & $10^{-5}$ & 0.01 & pow \\
\cline{2-6}
 & Equal rights & 0.045 & $10^{-6}$ & 0.01 & pow \\
\hline\hline
\multirow{6}{*}{1:5 (w5)} & Euthanasia & 0.025 & $3.2\times10^{-5}$ & 0.005 & pow \\
\cline{2-6}
 & Fed. spend. on social security & 0.035 & $10^{-7}$ & 0.1 & exp \\
\cline{2-6}
 & Fed. spend. on welfare & 0.025 & $3.2\times10^{-5}$ & 0.005 & pow \\
\cline{2-6}
 & Job guarantee & 0.6 & 0.5 & 0.6 & pow \\
\cline{2-6}
 & Marijuana & 0.005 & $10^{-5}$ & 0.01 & pow \\
\cline{2-6}
 & Equal rights & 0.045 & $10^{-6}$ & 0.01 & pow \\
\hline\hline
\multirow{6}{*}{1:10 (w10)} & Euthanasia & 0.01 & 0.05 & 0.02 & pow \\
\cline{2-6}
 & Fed. spend. on social security & 0.3 & $10^{-6}$ & 0.25 & exp \\
\cline{2-6}
 & Fed. spend. on welfare & 0.3 & $10^{-6}$ & 0.25 & exp \\
\cline{2-6}
 & Job guarantee & 0.6 & 0.5 & 0.6 & pow \\
\cline{2-6}
 & Marijuana & 0.3 & $10^{-6}$ & 0.25 & exp \\
\cline{2-6}
 & Equal rights & 0.045 & $10^{-6}$ & 0.01 & pow \\
\hline\hline
\multirow{6}{*}{1:20 (w20)} & Euthanasia & 0.01 & 0.05 & 0.02 & pow \\
\cline{2-6}
 & Fed. spend. on social security & 0.3 & $10^{-6}$ & 0.25 & exp \\
\cline{2-6}
 & Fed. spend. on welfare & 0.3 & $10^{-6}$ & 0.25 & exp \\
\cline{2-6}
 & Job guarantee & 0.6 & 0.5 & 0.6 & pow \\
\cline{2-6}
 & Marijuana & 0.3 & $10^{-6}$ & 0.25 & exp \\
\cline{2-6}
 & Equal rights & 0.01 & 0.05 & 0.02 & pow \\
\hline\hline
\multirow{6}{*}{1:40 (w40)} & Euthanasia & 0.01 & 0.05 & 0.02 & pow \\
\cline{2-6}
 & Fed. spend. on social security & 0.3 & $10^{-6}$ & 0.25 & exp \\
\cline{2-6}
 & Fed. spend. on welfare & 0.3 & $10^{-6}$ & 0.25 & exp \\
\cline{2-6}
 & Job guarantee & 0.6 & 0.5 & 0.6 & pow \\
\cline{2-6}
 & Marijuana & 0.3 & $10^{-6}$ & 0.25 & exp \\
\cline{2-6}
 & Equal rights & 0.01 & 0.05 & 0.02 & pow \\
\hline\hline
\multirow{6}{*}{1:80 (w80)} & Euthanasia & 0.01 & 0.05 & 0.02 & pow \\
\cline{2-6}
 & Fed. spend. on social security & 0.3 & $10^{-6}$ & 0.25 & exp \\
\cline{2-6}
 & Fed. spend. on welfare & 0.3 & $10^{-6}$ & 0.25 & exp \\
\cline{2-6}
 & Job guarantee & 0.6 & 0.5 & 0.6 & pow \\
\cline{2-6}
 & Marijuana & 0.3 & $10^{-6}$ & 0.25 & exp \\
\cline{2-6}
 & Equal rights & 0.01 & 0.05 & 0.02 & pow \\
\hline
\end{tabular}
\end{table}

\begin{table}[!ht]
\caption{The best parameters - SJBO}
\label{tab:best_params_sjbo}
\setlength{\extrarowheight}{2pt}
\centering
\begin{tabular}{|c|c|c|c|c|c|c|c|c|}
\hline
\multirow{2}{*}{\textbf{Sample weights}} & \multirow{2}{*}{\textbf{Topics}} & \multicolumn{7}{c|}{\textbf{The best parameters}} \\
\cline{3-9}
 &  & $\epsilon$ & $a$ & $r$ & $h$ & $\rho$ & $\lambda$ & $\tau$ \\
\hline
\multirow{6}{*}{1:1 (w1)} & Euthanasia & 0.6 & 0.25 & 0.85 & 0.25 & 0.25 & 0.85 & 1.2 \\
\cline{2-9}
 & Fed. spend. on social security & 0.6 & 0.25 & 0.85 & 0.25 & 0.25 & 0.85 & 1.2 \\
\cline{2-9}
 & Fed. spend. on welfare & 0.6 & 0.25 & 0.85 & 0.25 & 0.25 & 0.85 & 1.2 \\
\cline{2-9}
 & Job guarantee & 0.8 & 0.05 & 0.45 & 0.55 & 0.55 & 0.55 & 1.6 \\
\cline{2-9}
 & Marijuana & 0.6 & 0.25 & 0.85 & 0.25 & 0.25 & 0.85 & 1.2 \\
\cline{2-9}
 & Equal rights & 0.8 & 0.1 & 0.2 & 0.25 & 0.3 & 0.9 & 1.4 \\
\hline\hline
\multirow{6}{*}{1:3 (w3)} & Euthanasia & 1.3 & 0.01 & 0.6 & 0.2 & 0.65 & 0.9 & 1.8 \\
\cline{2-9}
 & Fed. spend. on social security & 1.35 & 0.01 & 0.15 & 0.1 & 0.35 & 0.8 & 1.75 \\
\cline{2-9}
 & Fed. spend. on welfare & 0.8 & 0.05 & 0.45 & 0.55 & 0.55 & 0.55 & 1.6 \\
\cline{2-9}
 & Job guarantee & 1.55 & 0.01 & 0.5 & 0.25 & 0.4 & 0.85 & 1.95 \\
\cline{2-9}
 & Marijuana & 0.6 & 0.45 & 0.9 & 0.2 & 0.5 & 0.55 & 1.3 \\
\cline{2-9}
 & Equal rights & 0.9 & 0.25 & 0.45 & 0.45 & 0.65 & 0.7 & 1.7 \\
\hline\hline
\multirow{6}{*}{1:5 (w5)} & Euthanasia & 1.3 & 0.01 & 0.6 & 0.2 & 0.65 & 0.9 & 1.8 \\
\cline{2-9}
 & Fed. spend. on social security & 1.05 & 0.05 & 0.25 & 0.35 & 0.8 & 0.75 & 1.75 \\
\cline{2-9}
 & Fed. spend. on welfare & 1.4 & 0.01 & 0.15 & 0.15 & 0.75 & 0.7 & 2.0 \\
\cline{2-9}
 & Job guarantee & 1.55 & 0.01 & 0.5 & 0.25 & 0.4 & 0.85 & 1.95 \\
\cline{2-9}
 & Marijuana & 1.7 & 0.01 & 0.15 & 0.3 & 0.6 & 0.9 & 1.9 \\
\cline{2-9}
 & Equal rights & 1.05 & 0.05 & 0.25 & 0.35 & 0.8 & 0.75 & 1.75 \\
\hline\hline
\multirow{6}{*}{1:10 (w10)} & Euthanasia & 1.8 & 0.01 & 0.7 & 0.25 & 0.75 & 0.8 & 2.0 \\
\cline{2-9}
 & Fed. spend. on social security & 1.05 & 0.05 & 0.25 & 0.35 & 0.8 & 0.75 & 1.75 \\
\cline{2-9}
 & Fed. spend. on welfare & 1.8 & 0.45 & 0.35 & 0.15 & 0.45 & 0.8 & 2.0 \\
\cline{2-9}
 & Job guarantee & 1.9 & 0.45 & 0.4 & 0.35 & 0.65 & 0.5 & 2.0 \\
\cline{2-9}
 & Marijuana & 1.9 & 0.45 & 0.4 & 0.35 & 0.65 & 0.5 & 2.0 \\
\cline{2-9}
 & Equal rights & 1.05 & 0.05 & 0.25 & 0.35 & 0.8 & 0.75 & 1.75 \\
\hline\hline
\multirow{6}{*}{1:20 (w20)} & Euthanasia & 1.8 & 0.01 & 0.7 & 0.25 & 0.75 & 0.8 & 2.0 \\
\cline{2-9}
 & Fed. spend. on social security & 1.65 & 0.1 & 0.65 & 0.45 & 0.8 & 0.75 & 2.0 \\
\cline{2-9}
 & Fed. spend. on welfare & 1.9 & 0.5 & 0.8 & 0.15 & 0.7 & 0.95 & 2.0 \\
\cline{2-9}
 & Job guarantee & 1.9 & 0.45 & 0.4 & 0.35 & 0.65 & 0.5 & 2.0 \\
\cline{2-9}
 & Marijuana & 1.9 & 0.45 & 0.4 & 0.35 & 0.65 & 0.5 & 2.0 \\
\cline{2-9}
 & Equal rights & 1.05 & 0.05 & 0.25 & 0.35 & 0.8 & 0.75 & 1.75 \\
\hline\hline
\multirow{6}{*}{1:40 (w40)} & Euthanasia & 1.8 & 0.01 & 0.7 & 0.25 & 0.75 & 0.8 & 2.0 \\
\cline{2-9}
 & Fed. spend. on social security & 1.65 & 0.1 & 0.65 & 0.45 & 0.8 & 0.75 & 2.0 \\
\cline{2-9}
 & Fed. spend. on welfare & 1.9 & 0.5 & 0.8 & 0.15 & 0.7 & 0.95 & 2.0 \\
\cline{2-9}
 & Job guarantee & 1.9 & 0.45 & 0.4 & 0.35 & 0.65 & 0.5 & 2.0 \\
\cline{2-9}
 & Marijuana & 1.9 & 0.45 & 0.4 & 0.35 & 0.65 & 0.5 & 2.0 \\
\cline{2-9}
 & Equal rights & 1.05 & 0.05 & 0.25 & 0.35 & 0.8 & 0.75 & 1.75 \\
\hline\hline
\multirow{6}{*}{1:80 (w80)} & Euthanasia & 1.8 & 0.01 & 0.7 & 0.25 & 0.75 & 0.8 & 2.0 \\
\cline{2-9}
 & Fed. spend. on social security & 1.65 & 0.1 & 0.65 & 0.45 & 0.8 & 0.75 & 2.0 \\
\cline{2-9}
 & Fed. spend. on welfare & 1.9 & 0.5 & 0.8 & 0.15 & 0.7 & 0.95 & 2.0 \\
\cline{2-9}
 & Job guarantee & 1.9 & 0.45 & 0.4 & 0.35 & 0.65 & 0.5 & 2.0 \\
\cline{2-9}
 & Marijuana & 1.9 & 0.45 & 0.4 & 0.35 & 0.65 & 0.5 & 2.0 \\
\cline{2-9}
 & Equal rights & 1.05 & 0.05 & 0.25 & 0.35 & 0.8 & 0.75 & 1.75 \\
\hline
\end{tabular}
\end{table}

\section{Correlation}
\label{app:correlation}
\begin{figure}[!ht]
	\begin{center}
        \includegraphics[scale=0.45]{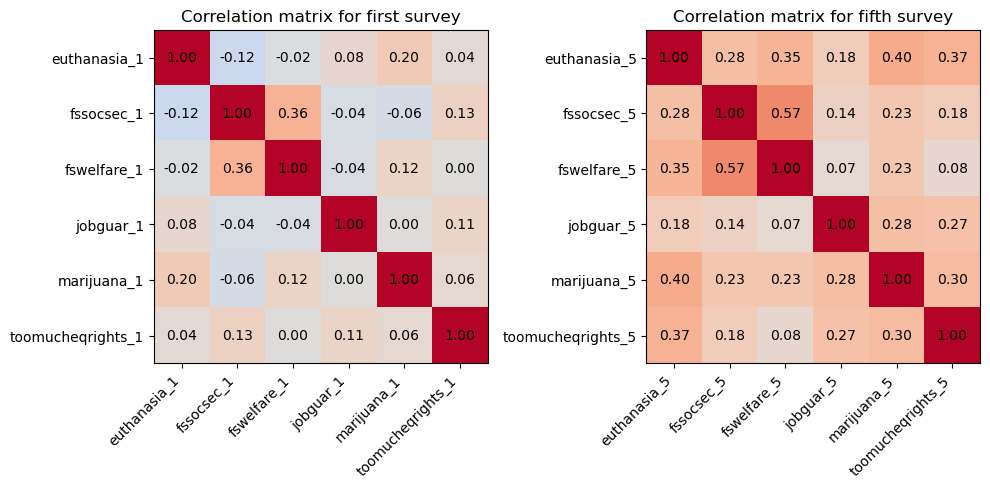}
        \caption{Correlation between question answers}
        \label{fig:correlation}
	\end{center}
\end{figure}

Figure~\ref{fig:correlation} shows the pairwise correlation between the answers to the six survey questions, for the first and the fifth survey separately. This complements the description of the dataset in Section\ref{subsec:dataset} of the main article. No strong correlation is present between individual topics, the overall level of correlation is higher in the fifth survey than in the first.

\section{Data split}
\begin{figure}[!ht]
	\begin{center}
        \includegraphics[scale=0.2]{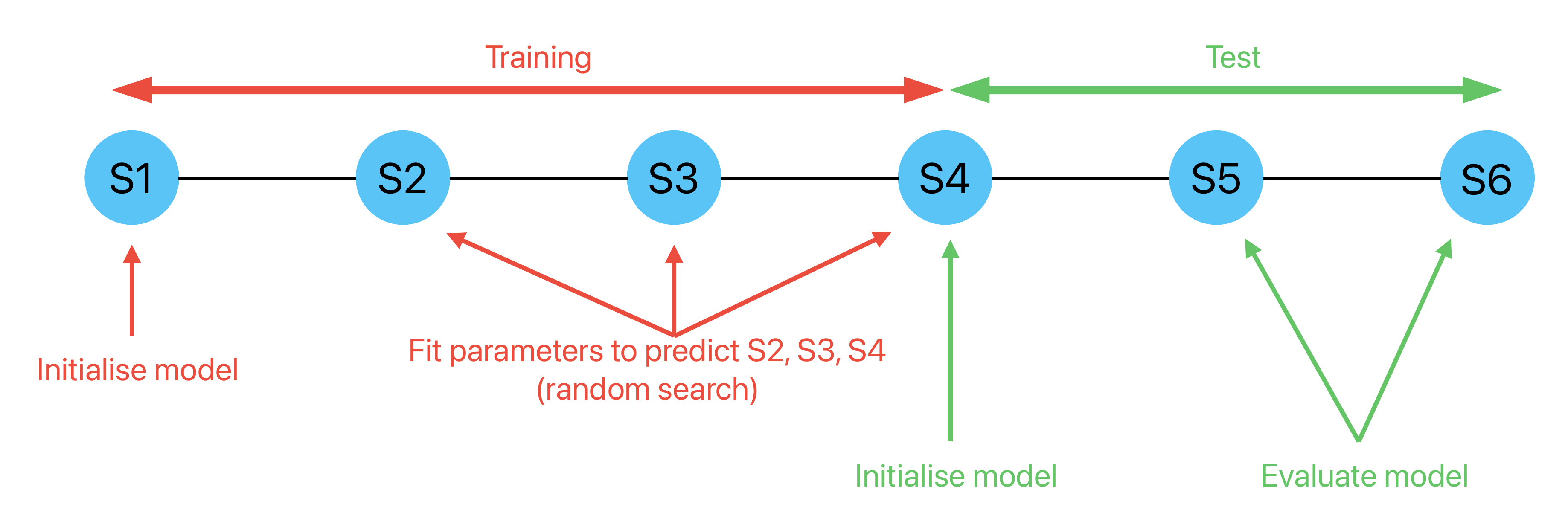}
        \caption{Data split strategy, illustrating the division of surveys 1--4 for parameter fitting, and surveys 5--6 for evaluation.}
        \label{fig:data_split}
	\end{center}
\end{figure}

Figure~\ref{fig:data_split} illustrates the split of the data into a training and a testing part used in Section~\ref{subsec:setting} of the main article. Surveys one to four form the training part, where the first survey initialises the agent states and the parameters are fitted by random search to reproduce surveys two, three, and four. In the testing part, survey four initialises the agent states, and surveys five and six are held out and used only as ground truth for the final evaluation.

\section{Sensitivity Analysis and Ablation}
\label{app:sensitivity}
\begin{figure}[!ht]
    \centering
    \subfloat[\label{fig:sens_euthanasia}]
    {\includegraphics[scale=0.34]{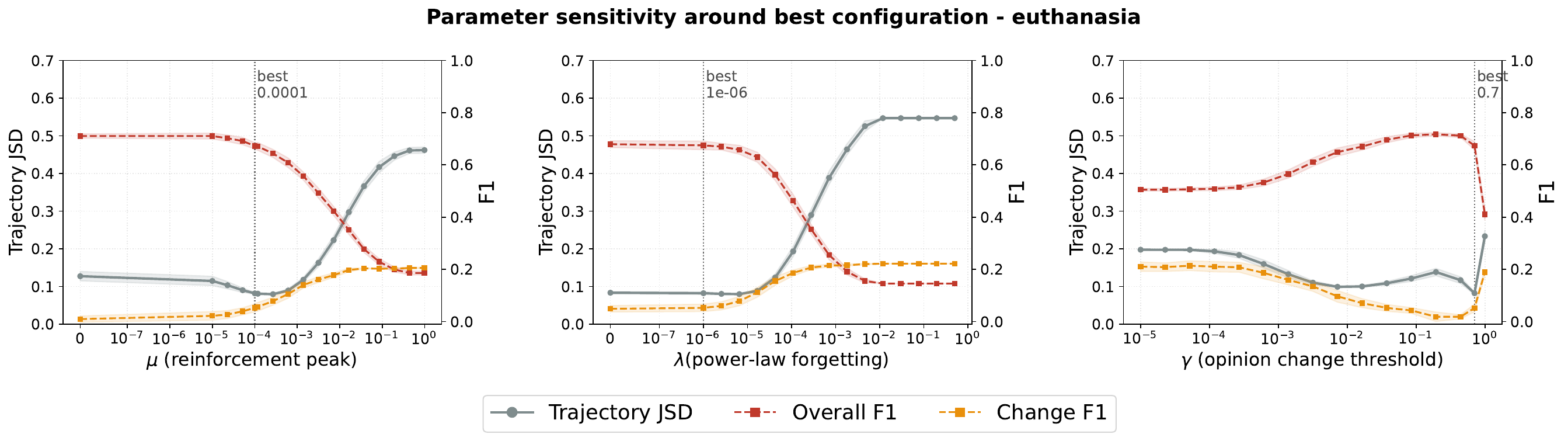}}
    \vspace{1em}
    \subfloat[\label{fig:sens_fswelfare}]
    {\includegraphics[scale=0.34]{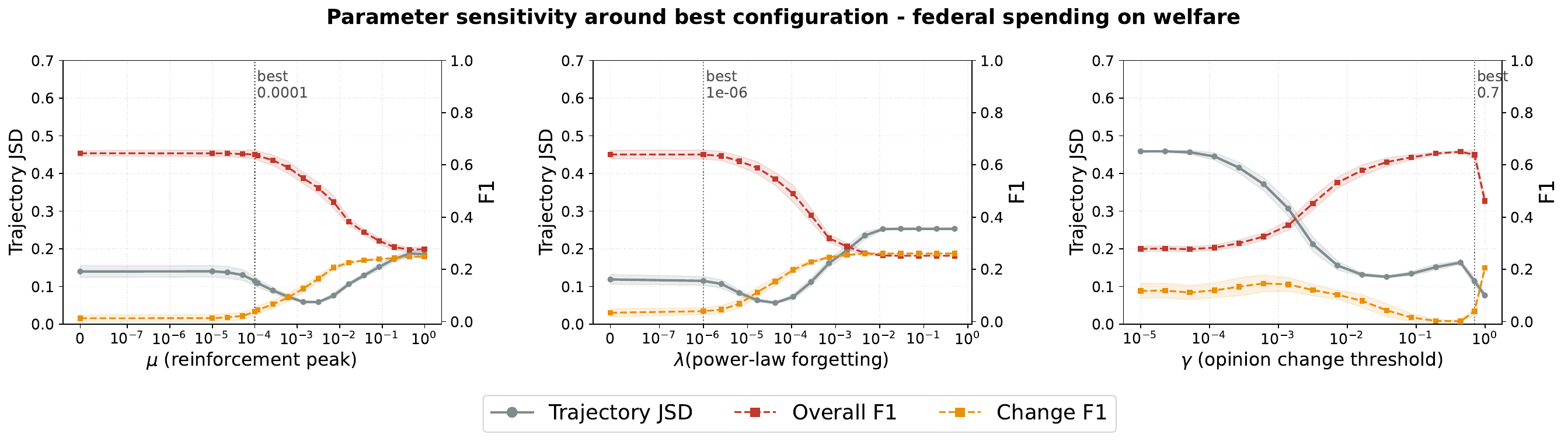}}
    \vspace{1em}
    \subfloat[\label{fig:sens_marijuana}]
    {\includegraphics[scale=0.34]{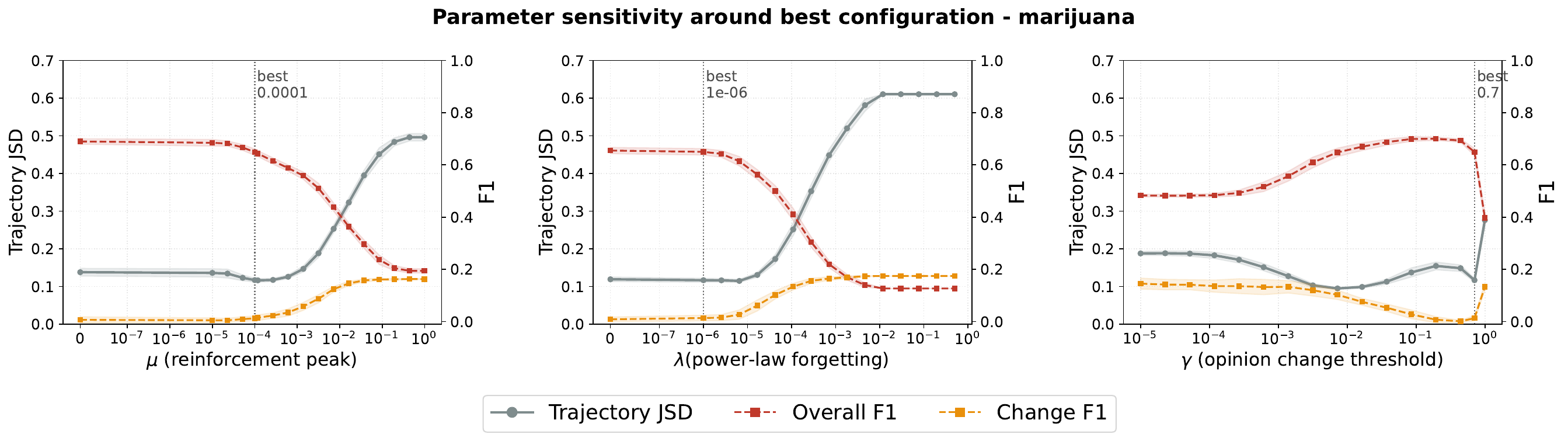}}
    \vspace{1em}
    \subfloat[\label{fig:sens_tomucheqrights}]
    {\includegraphics[scale=0.34]{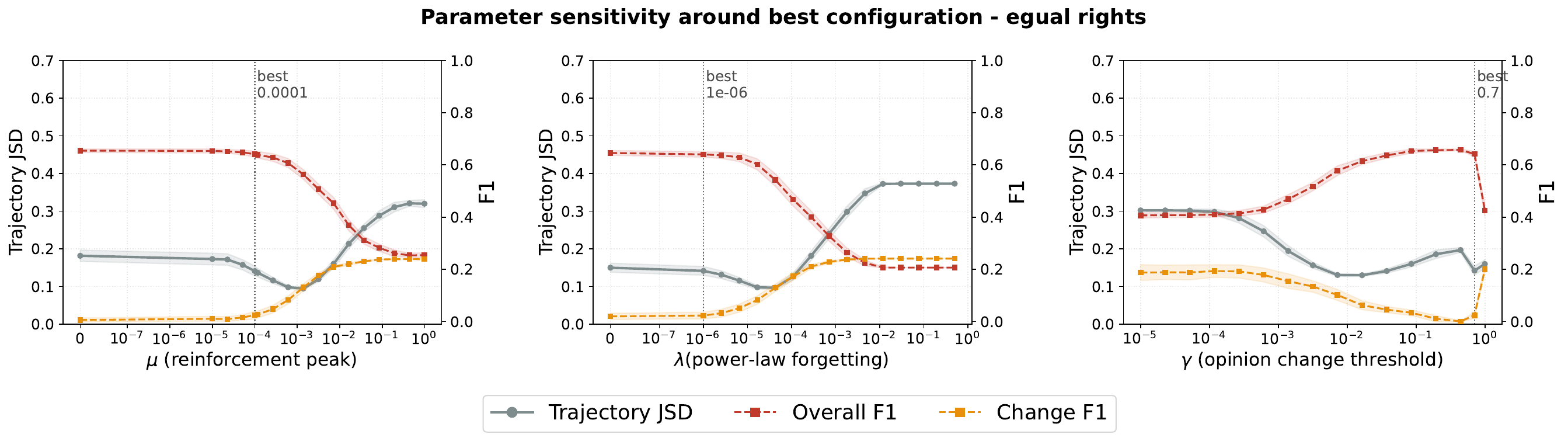}}
    \caption{Sensitivity and ablation analysis for selected topics, showing the effect of varying a single parameter ($\mu$, $\lambda$, or $\gamma$) while keeping others fixed.}
    \label{fig:sensitivity}
\end{figure}

Figure~\ref{fig:sensitivity} extends the sensitivity and ablation analysis of Section~\ref{subsec:sensitivity} of the main article to the remaining topics. For each topic, one parameter ($\mu$, $\lambda$, or $\gamma$) is varied around the best configuration while the others are held fixed. Setting $\mu=0$ disables the reinforcement mechanism, whereas setting $\lambda=0$ disables the forgetting mechanism, resulting in permanent memory. Federal spending on social security and job guarantee are shown in Section~\ref{subsec:sensitivity} of the main article; the topics reported here (euthanasia, federal spending on welfare, marijuana, and equal rights) exhibit similar behavior.

\end{document}